\documentclass[twocolumn,useAMS,usenatbib]{mn2e}
\usepackage{graphicx}
\usepackage{natbib}

\usepackage{bm}
\usepackage{amsfonts}
\usepackage{latexsym}
\usepackage[latin1]{inputenc}
\usepackage{amsmath}
\usepackage{epsfig}
\usepackage{amsbsy}
\usepackage{amssymb}

\newcommand{\n}{\mathrm n}
\newcommand{\p}{\mathrm p}

\newcommand{\x}{\mathrm x}
\newcommand{\y}{\mathrm y}
\newcommand{\ch}{\mathrm c}
\newcommand{\f}{\mathrm f}

\newcommand{\re}{\mathrm e}

\newcommand{\veps}{\varepsilon}

\def \nn  {\nonumber}

\def \eps{\epsilon}
\def \veps{\varepsilon}
\def\jnl@style{\it}
\def\aaref@jnl#1{{\jnl@style#1}}

\def\aaref@jnl#1{{\jnl@style#1}}

\def\aj{\aaref@jnl{AJ}}                   
\def\apj{\aaref@jnl{\rm ApJ}}                 
\def\apjl{\aaref@jnl{\rm ApJ}}                
\def\apjs{\aaref@jnl{ApJS}}               
\def\apss{\aaref@jnl{Ap\&SS}}             
\def\aap{\aaref@jnl{A$\&$A}}                
\def\aapr{\aaref@jnl{A$\&$A~Rev.}}          
\def\aaps{\aaref@jnl{A$\&$AS}}              
\def\mnras{\aaref@jnl{\rm MNRAS}}             
\def\prc{\aaref@jnl{\rm Phys.~Rev.~C}}        
\def\prd{\aaref@jnl{\rm Phys.~Rev.~D}}        
\def\prl{\aaref@jnl{\rm Phys.~Rev.~Lett.}}    
\def\qjras{\aaref@jnl{QJRAS}}             
\def\skytel{\aaref@jnl{S$\&$T}}             
\def\ssr{\aaref@jnl{Space~Sci.~Rev.}}     
\def\zap{\aaref@jnl{ZAp}}                 
\def\nat{\aaref@jnl{Nature}}              
\def\aplett{\aaref@jnl{Astrophys.~Lett.}} 
\def\apspr{\aaref@jnl{Astrophys.~Space~Phys.~Res.}} 
\def\physrep{\aaref@jnl{Phys.~Rep.}}      
\def\physscr{\aaref@jnl{Phys.~Scr}}       
\def\npa{\aaref@jnl{\rm Nucl. Phys. A}}                   
\def\plb{\aaref@jnl{\rm Phys. Lett. B}}                   

\title[QPOs in superfluid magnetars]{Quasi-periodic oscillations in
  superfluid, relativistic magnetars with nuclear pasta phases}

\author[A. Passamonti]
{Andrea Passamonti$^{1,2}$\thanks{E-mail:passamonti@ua.es}, Jos\'{e} A. Pons$^{1}$
\\ $^{1}$Department de Fisica Aplicada, Universitat d'Alacant, Ap. Correus 99, 03080 Alacant, Spain 
\\ $^{2}$INAF-Osservatorio Astronomico di Roma, via Frascati 44, I-00040, Monteporzio Catone (Roma), Italy}

\begin{document}

\date{\today}

\pagerange{\pageref{firstpage}--\pageref{lastpage}} \pubyear{}

\maketitle

\label{firstpage}


\begin{abstract}
We study the torsional magneto-elastic oscillations of relativistic
superfluid magnetars and explore the effects of a phase transition in
the crust-core interface (nuclear pasta) which results in a weaker
elastic response.  Exploring various models with different extension
of nuclear pasta phases, we find that the differences in the
oscillation spectrum present in purely elastic modes (weak magnetic
field), are smeared out with increasing strength of the magnetic
field. For magnetar conditions, the main characteristic and features
of models without nuclear pasta are preserved.  We find in general two
classes of magneto-elastic oscillations which exhibit a different
oscillation pattern.  For $B_p < 4 \times 10^{14}$G, the spectrum is
characterised by the turning points and edges of the continuum which
are mostly confined into the star's core, and have no constant phase.
Increasing the magnetic field, we find, in addition, several
magneto-elastic oscillations which reach the surface and have an
angular structure similar to crustal modes.  These global
magneto-elastic oscillations show a constant phase and become dominant
when $B_p > 5 \times 10^{14}$G.  We do not find any evidence of
fundamental pure crustal modes in the low frequency range (below 200
Hz) for $B_p \geq 10^{14}$G.
\end{abstract}

\begin{keywords}
methods: numerical -- stars: neutron -- stars: oscillations -- stars: magnetic fields
\end{keywords}

\section{Introduction} \label{sec:intro}

The first data set for applying Astereoseismology in neutron stars was
provided by magnetars.  The strong magnetic field of these stars,
$B>10^{14}$G, powers a rich X-ray activity of persistent and sporadic
emission~\citep{2008A&ARv..15..225M}. The most energetic events are
the giant flares which can radiate up to $10^{46}$~erg per second.  In
the tail of the three giant flares so far observed (SGR 0526-66, SGR
1900+14, SGR 1806-20), a series of Quasi Periodic Oscillations (QPOs)
with different duration and frequency was revealed by spectral
analyses~\citep{2005ApJ...628L..53I,2005ApJ...632L.111S,
  2006ApJ...637L.117W}.  The majority of them reside in the low
frequency band $18~\textrm{Hz} < \nu < 200$~Hz, but oscillations have
also been observed in the SGR 1806-20 at $625$~Hz and
$1840$~Hz. Recently, the analysis of the storm events in magnetars
have revealed a few QPOs also in intermediate flares
\citep{Hupp2014a, 2014ApJ...795..114H}, which occur more frequently than giant
flares but are less energetic.

Since their detection, it was immediately clear that QPOs could
originate from star oscillations and therefore used to explore the
properties of magnetar physics.  Initially, they were identified with
crustal shear modes~\citep{1998ApJ...498L..45D, 2005ApJ...628L..53I},
but it was soon pointed out that the strong magnetic field of
magnetars might dominate the global oscillations and significantly
influence the vibrations of the crust~\citep{2006MNRAS.368L..35L,
  2006MNRAS.371L..74G, 2007MNRAS.377..159L}.  In particular, torsional
magnetic oscillations of a purely poloidal magnetic field can produce
bands of continuum spectrum, which can rapidly absorb the crustal
modes that reside within them~\citep{2011MNRAS.410.1036V,
  2012MNRAS.420.3035V}.  In this theoretical model, the spectrum is
characterised by magnetically dominated modes, the turning points and edges of the
continuum bands, and the crustal modes whose frequency lies within the
continuum gaps. In particular, the interpretation of the $625$~Hz QPO
as a torsional shear overtone appeared difficult.  At this high
frequency the continuum bands should overlap and therefore quickly
damp this mode. Recently \citet*{2014ApJ...793..129H} have re-analysed
this QPO in the data and found that it can be consistent with either a
long-lived or an intermittent oscillation with decaying time $\sim
0.5s$.  In the latter case, the QPO must be re-excited during the
giant flare many times from a process at moment unknown.  The
continuum spectrum may be however an artefact of oversimplified
theoretical models, which consider magneto-elastic torsional
oscillations in a purely poloidal magnetic field.  Indeed, the
continuum could be reduced or destroyed by the interaction between the
torsional and poloidal modes via mixed poloidal-toroidal magnetic
field configurations \citep{2012MNRAS.423..811C} or by tangled
magnetic fields~\citep{2011MNRAS.410.1036V, 2015arXiv150301410L,
  2015PhRvD..92j4024S}.

Several works have studied the oscillations of magnetised neutron
stars with various degrees of model sophistication. With plane wave
approximation \citep{vHoven-Levin-2008, 2009MNRAS.396..894A}, in time domain
simulations~\citep{2007MNRAS.375..261S, 2008MNRAS.385L...5S,
  2009MNRAS.396.1441C, 2011MNRAS.410L..37G, 2011MNRAS.410.1036V, 2012MNRAS.420.3035V, 2012MNRAS.421.2054G,
  2013MNRAS.430.1811G} and as eigenvalue
problems~\citep{2008MNRAS.385.2069L, 2014ApJ...790...66A,
  2015MNRAS.449.3620A, 2016MNRAS.455.2228A}.  In particular,
\citet{2011MNRAS.414.3014C} were able to identify the QPOs with a set
of magnetic and crustal modes and a specific Equation of State (EoS)
without superfluid matter.

Mature neutron stars are however expected to contain superfluid and
superconducting constituents, which may strongly influence the
dynamics and the oscillation spectrum.  The effects of superfluidity
on the crustal modes have been studied by many authors
\citep{2009CQGra..26o5016S, 2012MNRAS.419..638P, 2013MNRAS.428L..21S},
while the effects on the non-axisymmetric magnetic oscillations were
addressed by~\citet{2013MNRAS.429..767P}. In superfluid magnetars, the
torsional magneto-elastic waves were recently studied in relativistic
\citep{2013PhRvL.111u1102G, 2016arXiv160507638G} and in Newtonian
stars \citep{2014MNRAS.438..156P}, which show a richer spectrum in
superfluid models.  Besides the magneto-elastic oscillations found in
non-superfluid stars, there are several waves which have an angular
structure similar to crustal oscillations. These waves are however
present both in the core and the crust and may have discrete
character~\citep{2013PhRvL.111u1102G, 2016arXiv160507638G}.

Another important aspect for the QPO interpretation is the wave
transmission from the star to the external magnetosphere.  In the
standard magnetar model, it is believed that the star's vibrations can
modulate the X-ray emission of a fireball anchored in the
magnetosphere near the star surface and thus produce the observed
QPOs. Progresses in this direction were recently made by
\citet{2014MNRAS.441.2676L} with a plane-wave analysis and by
\citet{2014MNRAS.443.1416G} with more elaborated models, which couple
the internal wave dynamics with the magnetosphere oscillations.

In this work, we study the torsional magneto-elastic oscillations of
magnetars.  In our model, we consider the effects of entrainment as
well as that of nuclear pasta phase on the oscillation spectrum.  This
new phase can be present as a transition region between the bottom of
the crust and the core \citep[e.g. see][and references
  therein]{2013PhRvC..88f5807S, 2015PhRvC..91f5802C}, where nuclei can
assume exotic shapes, like rods, plates, bubbles, etc. The elasticity
of this part of the crust can be quite different from regular cubic
lattices of ions \citep{1998PhLB..427....7P}, and its effects on the
crustal modes quite relevant \citep{2011MNRAS.417L..70S,
  2011MNRAS.418.2343G}.  These works have shown, in fact, that the
crustal modes have lower frequencies and a denser spectrum when the
pasta phase region is wider.  The possible identification of these
modes with the QPOs can therefore provide interesting results to
understand the physics of the crust.  However,
\citet{2011MNRAS.417L..70S} and \citet{2011MNRAS.418.2343G} focus on
the crustal modes and neglect the magnetic field, which, as described
before, can significantly modify the results.  In this work, we try to
clarify this issue and see whether the presence of nuclear pasta
phases changes the spectrum of the torsional magneto-elastic waves.

We present the formalism and the properties of our magnetar model in
Secs. \ref{sec:form} and \ref{sec:NSM}. The perturbation equations to
study the magneto-elastic torsional oscillations are given in
Sec. \ref{sec:pert} and in the Appendix, while the numerical framework
is described in Sec. \ref{sec:code}.  The results are presented in
Sec. \ref{sec:res} and the conclusions can be found in
Sec. \ref{sec:conc}.

\section{Formalism} \label{sec:form}

During the cooling of a neutron star, when the temperature drops below
the superfluid critical temperature, $T_{c} \simeq 10^{9}$K, the
neutrons of the core and the inner crust and the protons of the core
can become, respectively, superfluid and superconducting.  As a
result, the neutron interaction with the other particles is weaker and
completely different from the non-superfluid state. On the other hand,
protons and electrons in the core are so strongly coupled by
electromagnetic interaction that they can be considered as a single
comoving neutral fluid.  The dynamics is then naturally described by
two degrees of freedom, a gas of free superfluid neutrons and a
neutral mixture of charged particles, which for simplicity we will
call ``protons''.  In the inner crust, the second degree of freedom is
given by the heavy nuclei of the lattice.  To discern the various
components we use Roman letter x, y, \dots as constituent index.
Specifically, we denote with the letter $\n$ and $\p$ the superfluid
neutrons and the neutral mixture of the core, and with $\f$ and $\ch$
the free superfluid neutrons and the nuclei of the inner crust.  These
matter indices are not summed over when repeated.  We assume in this
work a strong superfluid regime, i.e. the star's temperature is well
below the neutron critical temperature $T_{\ch \n}$.

In general, superfluid gap models suggest that protons become
superconducting at $ T \approx 5 \times 10^{9}$K, when neutrons might
be still in a normal state.  The proton superfluid transition modifies
significantly the properties of the magnetic field, which now likely
reconfigures itself in an array of fluxtubes (Type II
superconductivity).  Although very interesting, we neglect in this
work the effects of superconductivity on the oscillation modes.
However, it is not still clear if the magnetic field in magnetars
exceeds the critical value ($ 10^{15} \textrm{G} - 10^{16}$G), above
which Type II superconductivity is destroyed.  Another possibility is
that the superconducting states could be limited to a shell near the
crust/core interface \citep{2014arXiv1403.2829S}.

We study the dynamics of this system with the relativistic two-fluid
model, based on the constrained variational approach developed by
Carter and collaborators \citep{1989LNM..1385....1C,
  1998NuPhB.531..478C, 2006CQGra..23.5367C}.  The fundamental
quantities of the constrained variational formalism are the master
function $\Lambda$ and the particle currents.  The particle fluxes are
defined by the following expression:
\begin{equation}
n_{\x}^{\alpha} \equiv n_{\x}   u^{\alpha}_{\x} \, ,  
\end{equation}
where $u^{\alpha}_{\x}$ is the velocity of the x fluid, and $n_{\x}$
is the particle density. It is determined by the normalization
condition, $n_{\x}^2 = - n_{\x }^{\alpha} n_\alpha^{\x}$.
The master function of a fluid star is in general a function of the
scalars which can be built from the number density currents, i.e.
$n_{\x}^2$ and $n_{\x\y}^2 = - n_{\x }^{\alpha} n_\alpha^{\y}$.  The
conjugate momenta arise naturally, in the variational approach, from
the definition
\begin{equation}
 \mu_{\alpha}^{\x} =  \left( \frac{\partial \Lambda}{ \partial n_{\x} ^{\alpha} } 
 \right) _{n_{\y}^{\alpha}}  \, .
\end{equation}
For superfluid relativistic stars with crust and magnetic field, the
master function as well as the derivation of the dynamical equations
have been described in detail by \citet{2006CQGra..23.5367C}
\citep[see also][for an application to non-magnetised
  stars]{2009CQGra..26o5016S}.  We do not provide here all the details
of the formalism but consider only the essential parts \citep[see][
  for a review]{2007LRR....10....1A}.

The dynamical equations are given by the particle number conservation
equations:
\begin{align}
 \nabla_{\alpha} n_{\x}^{\alpha} = 0 \, ,  \label{eq:pc}
\end{align}
 and by the momentum equations:
\begin{align}
& 2  n_{\f}^{\alpha} \nabla_{ \left[   \alpha \right.} \mu^{\f}_{   \left. \beta  \right]} = 0  \, ,   \label{eq:mf} \\ 
& 2  n_{\ch}^{\alpha} \nabla_{ \left[   \alpha \right.} \mu^{\ch}_{   \left. \beta  \right]} = f _{\beta}^{\ch} \, ,  \label{eq:mc}
\end{align}
where $f _{\beta}^{\ch}$ is the force density acting on the mixture of
charged particles $\ch$. The same equations are valid for the core,
provided the indices $\f$ and $\ch$ are, respectively, replaced with
$\n$ and $\p$.  The force density in the crust reads
\begin{align}
f _{\beta}^{\ch} = -  \nabla_{\alpha} \left( \pi^{\alpha}_{\beta}    + M^{\alpha}_{ \beta}  \right)     \, , 
\end{align}
where $\pi ^{\alpha \beta} $ is the stress-energy tensor of the solid
crust, while $M^{\alpha \beta}$ is the magnetic stress energy tensor.
In the core, the ``proton'' fluid feels only the magnetic force:
\begin{align}
f _{\beta}^{\p} = - \nabla_{\alpha}  M^{\alpha}_{ \beta}       \, .
\end{align}

In ideal MHD, the magnetic energy-momentum tensor is given by
\begin{equation}
 M^{\alpha \beta } =  \frac{1}{4 \pi}  \left( B^2 u^{\alpha}_{\x} u^{\beta}_{\x} 
+ \frac{B^2}{2}  g^{\alpha \beta } - B^{\alpha} B^{\beta} \right) \, ,
\end{equation}
where $\x=\p$ in the core and $\x=\ch$ in the crust.

The stress energy tensor for an isotropic crust can be derived with
the formalism introduced by \cite{1972RSPSA.331...57C} \citep[see
  also][for more general cases]{2003CQGra..20.3613K}.  The first step
is to define the stress energy tensor in terms of the shear modulus
$\check \mu$ and the shear tensor $\Sigma ^{\alpha \beta }$:
\begin{equation}
 \pi^{\alpha \beta } =  - 2 \check \mu \Sigma ^{\alpha \beta } \, .  
\end{equation}
Secondly, the shear tensor is defined by the following equation:
\begin{equation}
\mathcal{L}_{u} \Sigma ^{\alpha \beta } = \sigma ^{\alpha \beta } \, , 
\end{equation}
where $\mathcal{L}_{u}$ is the Lie derivative with respect to
$u^{\alpha}$. The stress tensor is
\begin{equation}
 \sigma^{\alpha \beta } =  \frac{1}{2} \left( u^{\alpha}_{\ch; \nu} \bot^{\nu \beta }  
 + u^{\beta}_{\ch; \nu} \bot ^{\nu \alpha } \right) 
 - \frac{1}{3} \bot ^{\alpha \beta } u^{\nu}_{\ch ; \nu} \, ,   
\end{equation}
and the projector tensor, $\bot^{\alpha \beta }$, is defined as follows
\begin{equation}
\bot^{\alpha \beta }  = g^{\alpha \beta }  + u_{\ch}^{\alpha }  u_{\ch}^{\beta }  \, . 
\end{equation}

To close the system of equations (\ref{eq:pc})-(\ref{eq:mc}) we need
an evolution equation for the magnetic field, which is given by the relativistic
induction equation. In ideal MHD it reads
\begin{equation}
\nabla _{\alpha} \left(  u_{\ch}^{\alpha} B^{\beta }  + u_{\ch}^{\beta} B^{\alpha } \right) = 0 \,  . \label{eq:ind}
\end{equation}
The usual replacement of $u_{\ch}^{\alpha}$ with $u_{\p}^{\alpha}$  must be done in the core.

\section{Neutron star model} \label{sec:NSM}

In this section, we describe the properties of our neutron star model,
which represents a relativistic superfluid magnetar with a poloidal
magnetic field and a realistic equation of state. The strong magnetic
field and the slow rotation of magnetars have a negligible effect on
the global structure of the star. These stars are therefore well
described by a spherically symmetric spacetime.  Considering a star
with an unstrained crust, the spacetime is given by the following line
element:
\begin{equation}
ds^2 = - e^{2 \nu } dt^2 + e^{2 \lambda} dr^2 + r^2 d\theta ^2 + r^2 \sin \theta ^2 d \phi^2 \, ,
\end{equation}
where $\nu$ and $\lambda$ are functions only of the radial coordinate.
They are determined by solving the Tolman-Oppenheimer-Volkoff
equations with a specific EoS. The background 4-velocity and the conjugate momenta, 
respectively, read
\begin{equation}
u^{\alpha} = \left( e^{-\nu} , \vec{0} \right)  \, , \qquad 
\mu^{\x}_{\alpha} = \left( -\mu^{\x} e^{\nu} , \vec{0} \right) \, , 
\end{equation}
where $\mu^{\x}$ is the chemical potential of the x fluid. 

\subsection{Equation of State} \label{sec:EoS}
                                 
We use the Douchin-Haensel EoS, which describes the state of matter
both in the core and the crust of a neutron star
~\citep{2001A&A...380..151D}. It is based on a Skyrme-type energy
density functional (SLy model).  Our baseline neutron star model has
$M=1.4 M_{\odot}$ and $R=11.66$~km, while the crust-core transition
appears at density $\rho_{cc} = 1.285\times 10^{14} \textrm{g
  cm}^{-3}$, which corresponds to a radial position $R_{cc} =
10.79$~km.  It is important to notice that the entrainment properties
illustrated in Sec. \ref{sec:entr} have not been derived for the DH
EoS.  Similar concerns apply to the pasta phases which are not present
in this EoS \citep{2000PhLB..485..107D}.  We use the DH EoS because it
provides all necessary inputs to model our superfluid neutron star and
to facilitate the comparison with previous works.  In any case, the
qualitative nature of our results should not change dramatically with
other EoSs which we plan to explore in future work \citep[see for
  instance][]{2012JPhCS.342a2003F, 2013A&A...560A..48P,
  2015A&A...584A.103S}.

\subsection{Background magnetic field} \label{sec:back}

The magnetic field in our model has a poloidal geometry and it is
determined by solving the Grad-Shafranov equation.  In a multi-fluid
system, the electric currents can be quite different from a single
fluid case.  For instance the superfluid neutrons do not contribute to
the inertia and in the crust only the free electrons can be able to
produce electric currents.  To simplify our approach we neglect the
effect of multi-fluid physics on the background magnetic field
configuration.  We therefore solve the Grad-Shafranov as in a single
fluid star and avoid unnecessary complexity at this point.
In fact, the uncertainty introduced by this assumption is smaller than
other approximations made in this kind of studies.  For instance, it
is well known that purely poloidal fields are unstable and cannot be
the final magnetic field configuration in a mature neutron star.
Furthermore, by adopting a single fluid magnetic field solution we can
much easily compare our results with the literature.

In the coordinate basis, the magnetic field components of a dipolar
field can be defined as follows
 \begin{equation}
B^r           = \re ^{-\lambda}  \frac{2 \cos \theta }{r^2}  a_1 \, , \qquad 
B^{\theta} = -{\re ^{-\lambda} }  \frac{\sin \theta }{ r^2}  \frac{d a_1 }{dr } \, , 
\end{equation}                         
where $a_1(r)$ is a solution of  the Grad-Shafranov equation:
\begin{equation}
 \frac{d^2 a_1 }{dr ^2 } +  \frac{d }{dr} \hspace{-0.08cm} \left(  \nu -  \lambda  \right) 
\frac{d a_1 }{dr } - \frac{ 2 }{r^2} \re ^{2 \lambda } a_1 
=  - 4 \mathrm{ \pi } c_0 r^2    \left( \veps + p \right) \re ^{2 \lambda }  \label{eq:GS} \, .
\end{equation}
The quantities $\veps$ and $p$ are, respectively, the mass-energy and
the pressure, while $c_0$ is a constant. To solve equation
(\ref{eq:GS}) we must specify the boundary conditions at both the
origin and the surface. Regularity of the solution at the origin
requires $a_1 = \alpha_0 r^2$ (when $r\to0$), where $\alpha_0$ is a
constant.  At the surface, the solution must be matched with an
external magnetic field, which we consider a dipolar field in the
vacuum. The condition at the surface is therefore:
\begin{equation}
 a_1= - \frac{3 \mu_0}{ 8 M^3} r^2 \left[  \ln \left( 1- \frac{2 M }{r} \right) + \frac{2 M }{r}
+ \frac{2 M^2 }{r^2}   \right]  \, , 
  \label{eq:a1ext}
\end{equation}
where $M$ is the star's mass and $\mu_0$ the magnetic dipole moment in
geometric units.  The two constants $\alpha_0$ and $c_0$ can be
determined from the numerical integration in order to satisfy the
boundary conditions. The solution is determined up to the arbitrary
multiplicative factor $\mu_0$, which is fixed by the value of the
poloidal magnetic field at the magnetic pole $B_{p}$. In our model,
the average magnetic field of the star is roughly $\langle B \rangle
\simeq 1.45 B_p$.

\subsection{Shear modulus} 

For the shear modulus in the crust we use the low temperature limit of
the formula derived by \citet{1991ApJ...375..679S}:
\begin{equation}
\check \mu = 0.1194 \left( \frac{4 \pi}{3}  \right) ^{1/3} n_i ^{4/3}  \left( Z e \right)^2 \, , \label{eq:sm}
\end{equation}
where $n_i$ is the ion number density and $Z e$ the ion
charge. Equation~(\ref{eq:sm}) assumes a body-centered-cubic
crystalline lattice, while at the bottom layers of the crust there
might be, in some EoS, stable configurations of ``pasta phases''.  In
these regions, the strong and Coulomb interaction can modify the shape
of the nuclei, which can adopt non-spherical shapes.  These exotic
phases can appear when the density is $\rho_0 \lesssim \rho \lesssim
0.1 \rho_0$, where $\rho_0 = 2.7\times 10^{14}~\textrm{g cm}^{-3}$ is
the nuclear saturation density~\citep{2013PhRvC..88f5807S}.  It is
worth noticing that if the various shapes which form the nuclear pasta
have regular structures, the shear modulus may be
anisotropic. However, if these structures are disordered, on average,
the anisotropy degree can be partially reduced
\citep{2013PhRvC..88f5807S, Horowitz-2015}.

\begin{figure}
\begin{center}
\includegraphics[height=75mm]{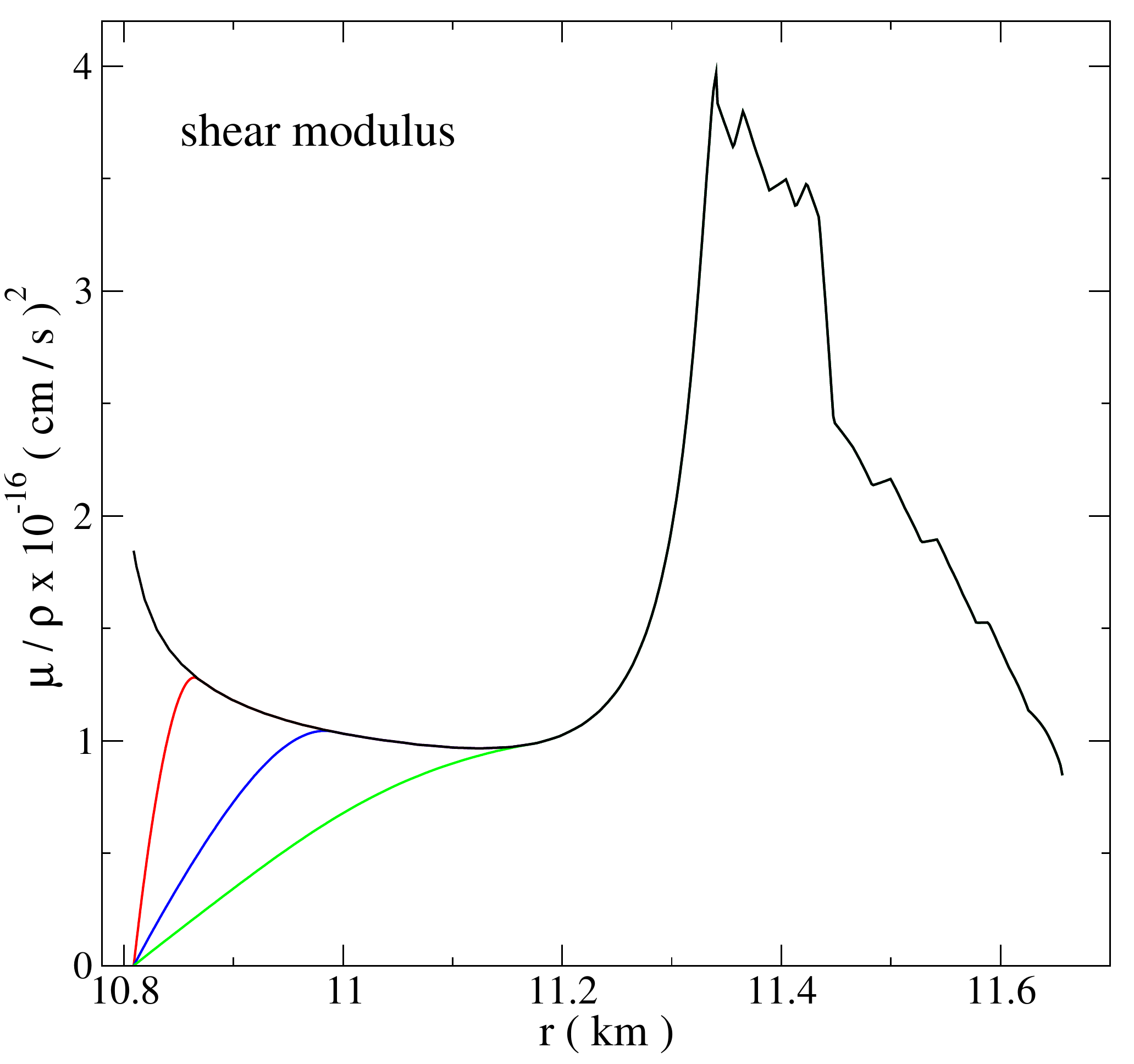} 
\caption{ Shear modulus ($\mu/\rho$) of our stellar model with
  different pasta phase transition density $\rho_{ph}$ . The black
  line describe a star without pasta phase, while the other three
  curves show model with decreasing $\rho_{ph}$. More precisely, the
  red line corresponds to $\rho_{ph} = 10^{14}$~g cm$^{-3}$, the blue
  line to $\rho_{ph} = 5\times10^{13}$~g cm$^{-3}$ and the green lines
  represent the $\rho_{ph} =10^{13}$~g cm$^{-3}$ case.
\label{fig:mu}}
\end{center}
\end{figure}

The elastic reaction of this exotic phases is still unknown, even if
some preliminary work is present in the literature \citep[see][for a
  review]{2008LRR....11...10C}, but in any case the rigidity of the
crust is expected to decrease.  To model the shear modulus between a
``normal'' solid crust, which is described by equation (\ref{eq:sm}),
and the core, where $\check \mu = 0$, we use a function which smoothly
joins these to regions given by
 \begin{equation}
\check \mu = c_1 \left( \rho - \rho_{cc} \right) \left( \rho - c_2
\right) \, , \label{eq:mupasta}
\end{equation}
where $\rho_{cc}$ is the density at the crust/core interface, and
$c_1$ and $c_2$ are two interpolation constants. Eq. \ref{eq:mupasta}
describes the shear modulus in the density range $\rho_{ph}\leq \rho
\leq \rho_{cc}$ where $\rho_{ph}$ is the pasta phase transition
density. In this work we consider this transition in the range $
10^{13} \textrm{g cm}^{-3} \lesssim \rho_{ph} \lesssim 10^{14}
\textrm{g cm}^{-3}$ (see Fig. \ref{fig:mu}).

The only difference between Eq. \ref{eq:mupasta} and the expression used
by~\citet{2011MNRAS.417L..70S} is in the term given by the first
bracket, which was originally squared, i.e.  $\left( \rho - \rho_{cc}
\right) ^2$. We have first tried the same expression, but we found
some problems of convergence for the eigenfrequencies of the shear
mode overtones. This problem is removed by using equation
(\ref{eq:mupasta}), which goes linearly to zero when $\rho$ gets close
to $\rho_{cc}$ (see Fig. \ref{fig:mu}).

\subsection{Entrainment} \label{sec:entr}

In superfluid stars the interaction between the various particle
species changes with respect to normal dissipative
processes. Superfluid neutrons and protons can interact via the mutual
friction which is a dissipative processes mediated by superfluid
vortices, which depends on the star's rotation rate.  Magnetars are
slowly rotating stars, thus this effect is likely negligible in the
magneto-elastic oscillations.  There is also another non-dissipative
process, called entrainment, that couples the dynamics of various
constituents of a superfluid system.  The entrainment in the core of a
neutron star is driven by the strong interaction force between the
nucleons and is typically weak. In the inner crust, superfluid
neutrons interact with the nuclei lattice via Bragg scattering and the
entrainment can be quite strong \citep[see][and reference
  therein]{2012PhRvC..85c5801C}.  A result of entrainment is that, in
a superfluid, the constituent momentum is not aligned with the
velocity, but contains also a contribution of the other particle
specie.  Normally, the entrainment is described by a quantity
$\veps_\x$ which is strictly related to the effective mass,
$m_{\x}^{\star}$, through the following relation:
\begin{equation}
\veps_{\x} = 1 - \frac{m_{\x}^{\star}}{m_{\x}} \, .  \label{eq:veps}
\end{equation}

In our model, we consider the entrainment profile provided in
\citet{2008MNRAS.388..737C} for the core, and
\citet{2005NuPhA.747..109C, 2006NuPhA.773..263C, 2012PhRvC..85c5801C}
for the crust. In practise, we use the same mathematical structure
described in \citet{2014MNRAS.438..156P}, but applied to a tabulated
EoS.  The maximum effect of the entrainment is expected in the inner
crust where the effective mass of neutrons can be as large as
$m_{\f}^{\star} \simeq 14 m_{\f}$.

\section{Perturbation Equations} \label{sec:pert}

In this section, we summarize the perturbation equations used to study
axisymmetric axial (torsional) oscillations of relativistic superfluid
magnetars. We use the Cowling approximation, i.e. we neglect the
perturbations of the space-time.  It provides accurate results for the
crustal and Alfv\'en torsional modes.

In a non-rotating axisymmetric star, the torsional modes satisfy the
number conservation equation by definition, $ \nabla_{\alpha } \delta
n_{\x}^{\alpha } = 0$.  The perturbed momentum conservation
equations~(\ref{eq:mf})-(\ref{eq:mc}) read
\begin{align}
&  2   \delta \left(  n_{\f}^{\alpha} \nabla_{ \left[   \alpha \right.} \mu^{\f}_{   \left. \beta  \right]}   \right) = 0 \, , \\
& 2  \delta \left( n_{\ch}^{\alpha} \nabla_{ \left[   \alpha \right.} \mu^{\ch}_{   \left. \beta  \right]} \right) = \delta f_{\beta}^{ \ch}  \, ,   
\end{align}
which, using the symmetry of the problem, leads to the following equations:
\begin{align}
&  \partial _t \delta \mu_{\phi}^{\f}  = 0 \, ,  \label{eq:dmuf} \\
&  n_{\ch} \, \re^{-\nu}  \, \partial _t  \delta  \mu^{\ch}_{ \phi}  = \delta f_{\phi}^{ \ch} \, . \label{eq:dmuc} 
\end{align}
Here, we have used the indices $\ch$ and $\f$ for the crust
constituents, but equations (\ref{eq:dmuf})-(\ref{eq:dmuc}) are
formally valid also for the core.  The conjugate momenta are related
to the velocity perturbations as follows:
\begin{align}
\delta \mu^{\f}_{\phi } =  \mathcal{K}_{\f  \f} n_{\f} \, \delta u^{\f}_{\phi} + \mathcal{K}_{\f  \ch} n_{\ch} \, \delta u^{\ch}_{\phi}  \,  , \\
\delta \mu^{\ch}_{\phi } =  \mathcal{K}_{\ch  \f} n_{\f} \, \delta u^{\f}_{\phi} + \mathcal{K}_{\ch  \ch} n_{\ch} \, \delta u^{\ch}_{\phi}   \, ,
\end{align}
where we have used the components of the entrainment matrix:
\begin{align}
& \mathcal{K}_{\f  \f}  = -2 \frac{\partial \Lambda}{ \partial n_{\f}^2}  \, , \\
&   \mathcal{K}_{\ch  \ch}   = -2 \frac{\partial \Lambda}{ \partial n_{\ch}^2}  \, , \\
& \mathcal{K}_{\f  \ch}  = \mathcal{K}_{\ch  \f}  = - \frac{\partial \Lambda}{ \partial n_{\ch \f}^2} \, .
\end{align}
Up to second order corrections in the relative velocity between the
$\ch$ and $\f$ fluids, the entrainment matrix is given by
\citep{2009CQGra..26o5016S}:
\begin{align}
& \mathcal{K}_{\f  \f}  =  \frac{1}{n n_{\f}} \left[  \left( \veps +p \right) - \rho \epsilon _{\f}   \right]   \, , \\
& \mathcal{K}_{\ch  \ch}   =  \frac{1}{n n_{\ch}} \left[  \left( \veps +p \right) - \frac{n_{\f}}{n_{\ch}} \rho \epsilon _{\f}   \right]  \, , \\
& \mathcal{K}_{\f  \ch}  = \mathcal{K}_{\ch  \f}  =  \frac{\rho \epsilon_{\f}}{n n_{\ch}}  \,  ,
\end{align}
where $n=n_{\ch} + n_{\f}$. 
The entrainment parameter $\veps_{\f}$ is related to the effective
mass of the free neutrons and has been defined in
equation~(\ref{eq:veps}).

The torsional oscillations in a spherical star with magnetic field can
be studied with a single wave-like equation for the conglomerate of
charged components.  From equation~(\ref{eq:dmuf}) we can write
\begin{equation}
 \partial _t \delta u^{\f}_{\phi }   = - \frac{ \mathcal{K}_{\f  \ch} }{\mathcal{K}_{\f  \f}}  \frac{ n_{\ch} }{ n_{\f}} \, \partial_t \delta u^{\ch}_{\phi} \, ,
\end{equation}
which  inserted in equation (\ref{eq:dmuc}) provides
\begin{equation}
\chi \left( \veps + p \right)  e^{-2\nu} \,  \frac{\partial^2 \xi_{\phi}^{\ch} }{\partial t^2} = \delta f_{\phi} ^{\ch} \, . 
\end{equation}
Here, we have defined the Lagrangian displacement $\xi_{\ch}^{\alpha}$ for the $\ch$ fluid:
\begin{equation}
\delta u _{\ch}^{\alpha} = \mathcal{L} _{u} \xi_{\ch}^{\alpha} = \re^{-\nu} \xi_{\ch}^{\alpha}  \, , 
\end{equation}
and a quantity that accounts for the entrainment \citep{2009CQGra..26o5016S}:
\begin{equation}
\chi = \frac{ x_{\ch} \left( \veps +p \right) - \rho \eps_{\f} }{ \veps +p  - \rho \eps_{\f}} = 1 - x_{\f}  \frac{\veps +p }{\veps +p  - \rho \eps_{\f}}  \,  ,
\end{equation}
where $x_{\ch} = n_{\ch} / n$ and $x_{\f} = n_{\f} / n$.  For a zero
entrainment system, $\veps_{\f} = 0$, and $\chi = x_{\ch}$.  Note that
in the Newtonian limit, $\chi$ tends to $ x_{\ch} \,
\veps_{\star}^{-1}$ which is the entrainment parameter used in
\citet{2014MNRAS.438..156P}.

In the crust the linearised force reads:
\begin{equation}
\delta f_{\phi} ^{\ch} = - \nabla_{\alpha} \delta \pi_{\, \, \phi}^{ \alpha} - \nabla_{\alpha}  \delta M_{\, \,  \phi}^{ \alpha} \, ,
\end{equation}
while in the core only the magnetic force is present, 
\begin{equation}
\delta f_{\phi} ^{\p} =  - \nabla_{\alpha}  \delta M_{\, \,  \phi}^{ \alpha} \, .
\end{equation}
The equations for the perturbation of the  magnetic stress-energy tensor 
have been already presented in several works \citep[see
  for instance][]{2007MNRAS.375..261S}. In particular, the
perturbation of the magnetic field $\delta B^{\alpha}$, can be
expressed in terms of the Lagrangian displacement $\xi^{\alpha}$ by
using the linearised induction equation
\begin{equation}
\delta  \left[ \nabla _{\alpha} \left(  u^{\alpha}_{\ch} B^{\beta }  + u^{\beta}_{\ch} B^{\alpha } \right) \right] = 0 \, .
\end{equation}
Without repeating the same calculations we directly provide the final
linearised equation, written in the coordinate basis, to study
axisymmetric torsional oscillations in a magnetised superfluid star
with crust:
\begin{align}
\left[ \chi \left( \veps +p \right) +\frac{B^2}{4 \pi} \right]  \textrm{e}^{-2 \nu } \frac{ \partial^2 \xi_{\ch}^{\phi}}{\partial t^2}  & =   A_1 \frac{ \partial^2 \xi_{\ch}^{\phi}}{\partial r^2} 
 + A_2 \frac{ \partial^2 \xi_{\ch}^{\phi}}{\partial \theta^2} 
+  A_3 \frac{ \partial^2 \xi_{\ch}^{\phi}}{\partial r \partial \theta}  \nn \\ 
& +  A_4 \frac{ \partial \xi_{\ch}^{\phi}}{\partial r} 
+ A_5 \frac{ \partial \xi_{\ch}^{\phi}}{\partial \theta}   \, .
\label{eq:wv}
\end{align}
The coefficients $A_{k}$ depends on the background variables and are
given in the Appendix~\ref{sec:wvcoef}.  The core's protons obey the
same wave equation with the coefficients $A_{k}$ taken in the zero
shear modulus limit, $ \check \mu = 0$.

\subsection{Boundary conditions} \label{sec:bd}

The symmetry of the problem allows us to study the time evolution of
equation (\ref{eq:wv}) in the 2D-region $0\leq r \leq R$ and $0\leq
\theta \leq {\rm \pi} / 2$.  At the boundary of this domain we must
obviously impose appropriate conditions.

The regularity of the perturbation equation at the origin requires
that $\xi^\phi _\p = 0$ at $r=0$.  At the magnetic axis, $\theta=0$,
the Lagrangian displacement must satisfy, in the coordinate basis, the
following relation $ \partial_\theta \xi^\phi _\x = 0$.

The magneto-elastic waves can be symmetric or antisymmetric with
respect to the equator, $\theta = \pi/2$.  Symmetric perturbations
obey the condition $ \partial_\theta \xi^\phi _\x = 0 $, and
antisymmetric perturbations must vanish at the equator $ \xi^\phi _\x
= 0 $.  In this work, we focus on the antisymmetric perturbations
which have the right symmetry to move the surface footprints of the
magnetic field in opposite directions and produce oscillations in a
`twisted magnetosphere'. These twists are believed to modulate the
X-ray emission \citep{2014MNRAS.443.1416G} and produce QPOs.

At the star's surface we match our internal solution with an external
magnetosphere. If we do not allow current sheets on the star's
surface, the perturbation of the $\phi$-component of the magnetic
field is continuous.  This leads to $\partial_r \xi_{\ch}^{\phi} = 0 $
at $r=R$ \citep{2011MNRAS.410L..37G}.

Due to the presence of an elastic crust, we must also consider the
junction conditions at the crust/core interface, $r=R_{cc}$.  We
impose here the continuity of the Lagrangian displacement,
\begin{equation}
\xi_{\ch} ^{\phi} = \xi_{\p} ^{\phi}  \label{eq:bc1} \, , 
\end{equation}
and the continuity of the traction,
\begin{equation}
\delta t^{r \phi} = \left( \mu + \re ^{-2\lambda} \frac{B_r ^2}{4 \pi} \right) \frac{ \partial \xi_{\ch}^{\phi}}{\partial r } 
+ \re ^{2 \lambda} \frac{B^r B^{\theta}}{4 \pi}  \frac{ \partial \xi_{\ch} ^{\phi}}{\partial \theta }  \, .  \label{eq:trac}
\end{equation}
From equation (\ref{eq:bc1}) and assuming that the tangential
derivatives are continuous on $r=R_{cc}$ we obtain a condition for the
radial derivatives of the Lagrangian displacement:
\begin{equation}
\left( \re ^{2\lambda}   \mu + \frac{B_r ^2}{4 \pi} \right) \frac{ \partial \xi_{\ch}^{\phi}}{\partial r }   = 
\frac{B_r ^2}{4 \pi} \frac{ \partial \xi_{\p}^{\phi} }{\partial r } \, .
\end{equation}

\begin{figure}
\begin{center}
\includegraphics[trim = 0mm 0mm 0mm 7cm, clip, height=36mm]{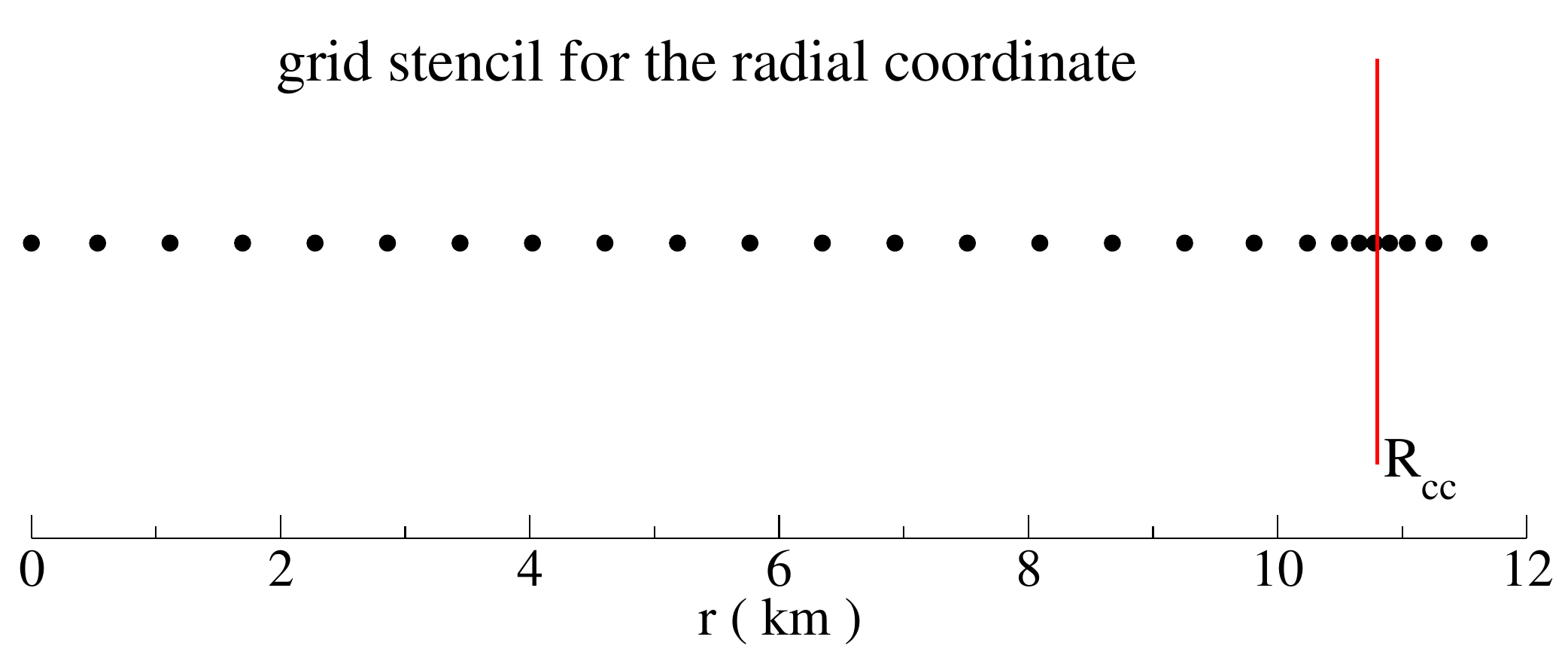}
\caption{This figure shows the grid stencil for the radial coordinate,
  which results from the coordinate transformation
  (\ref{eq:grid1}). The refinement of the mesh near the crust/core
  interface is evident. For clarity, we have shown here a grid with 26
  points, while in this work we use 200 points.
\label{fig:sten}}
\end{center}
\end{figure}

\section{Numerical method} \label{sec:code}

The numerical code to study the time evolutions of
equation~(\ref{eq:wv}) is based on the numerical framework developed
in \citet{2014MNRAS.438..156P}. The spatial partial derivatives are
approximated by using a second order finite difference scheme, while
the evolution in time is performed with an iterative explicit
Crank-Nicholson algorithm.  To stabilise the simulation we add also a
fourth order Oliger-Kreiss numerical dissipation, $\varepsilon_{D} D_4
\xi$, with a small dissipation coefficient $\veps_{D} \sim 10^{-5}$.
The numerical domain covers the 2D-region $0\leq r \leq R$ and $0\leq
\theta \leq {\rm \pi} / 2$, and the grid resolution used in this work
is 200 points in the radial coordinate and 64 points in the angular
coordinate.

The crust/core interface is the region where the core and crust
oscillations interact, and it is very important to have an adequate
grid resolution to describe the wave dynamics. Therefore, we introduce
the following transformation to refine the grid around $R_{cc}$:
\begin{equation}
\frac{dr}{dx} =  \beta_r   \,  ,  \label{eq:grid}
\end{equation} 
where
\begin{equation}
\beta_r  \equiv  1- a \exp \left[ {- \frac{1}{2} \left(\frac{r-r_0}{b}  \right)^2 }\right]  \, .   \label{eq:grid1} 
\end{equation} 
We use $a=0.8$, $b=0.5$~km and $r_0 = 10.8$~km, which is close to the
actual crust/core transition.  Considering an even spaced grid in the
x radial coordinate, equation~(\ref{eq:grid}) guarantees that the mesh
of the r coordinate is finer nearby the crust/core interface (see
Fig. \ref{fig:sten}).  The radius of our stellar model in the
x-coordinate is $R_x = 14.54$~km.

To evolve in time the perturbation equation $(\ref{eq:wv})$ in the new
grid, we replace the radial derivatives with the following expression:
\begin{equation}
\frac{\partial }{\partial r}  \to   \frac{1}{\beta_r}    \frac{\partial }{\partial x}   \label{eq:dergrid}  \,  ,
\end{equation} 
and the same transformation is applied also to the boundary
conditions.  We determine the properties of the magneto-elastic modes
by post-processing the results of the time simulations. First we get
the mode frequencies by using a Fast Fourier Transformation (FFT) of
the time evolved Lagrangian displacement $\xi^{\phi}_{\ch}$.  To
monitor the mode pattern we perform an effective FFT on the entire
grid by using a code developed by \citet{Stergioulas:2003ep}.  To
study the presence of the continuum spectrum we extract the FFT at
different positions inside the star.

\section{Results} \label{sec:res}

Several magnetar internal properties (composition, superfluidity,
extent of the nuclear pasta region) can influence the QPO spectrum. We
begin our discussion by reviewing the main characteristics of the QPOs
spectra. We first consider the crustal torsional modes in models
without magnetic field, then we study the oscillations in magnetised
stars without pasta phases, and finally address the full problem with
both magnetic field and pasta configurations.

With respect to superfluidity, the first order effect of the
decoupling of the superfluid neutrons from the charged components
leads to an increase of the crustal and Alfv\'en mode frequencies by a
factor $x_{\ch}^{-1/2}$. This can be seen from the definition of the
Alfv\'en and shear velocities, which in a two-fluid Newtonian system,
respectively, read
\begin{equation}
v_{A} = \frac{B}{\sqrt{4\pi\rho_{\ch}}}  \, ,  \qquad  v_{s} = \sqrt{\frac{\check \mu}{ \rho_{\ch}}} \, , 
\end{equation}
where the charged particle mass density now replaces the total density
appropriate for single fluid stars.  However, this frequency increase
can be mitigated in the inner crust by the strong entrainment (see
Sec. \ref{sec:cm}).  The influence of entrainment on the
magneto-elastic oscillations has been already studied in Newtonian
models by~\citet{2013MNRAS.429..767P, 2014MNRAS.438..156P} and
recently in the relativistic case by \citet{2016arXiv160507638G}. In
order not to repeat the same analysis, in this work we consider only
the single entrainment configuration which has been described in
Sec.~\ref{sec:entr}.

\begin{table}
\begin{center}
\caption{Frequencies of torsional crustal modes $^l t_n$ (in Hz) for
  non-magnetised neutron stars.  FD and TD correspond to frequencies
  determined with frequency and time domain approaches, respectively.
  The second and third columns show the results for a model without
  entrainment, while the fourth and fifth columns are calculated with
  the strong entrainment discussed in Sec.  \ref{sec:entr}.  The sixth
  column displays the mode frequencies for a non-superfluid (single
  fluid) star.
\label{tab1}}
\begin{tabular}{   c c c c c c  }
\hline
  Mode       & FD            &  TD           &         FD     &  TD            & FD          \\
             & no entr.      &   no entr.    &    with entr.  &  with entr.    & 1-fluid     \\
\hline 
$ ^{2}t_{0} $ &     51.1     &     51.4       &     27.4       &    27.7       &    25.1       \\
$ ^{3}t_{0} $ &     80.7     &     81.5       & 	  43.2        &   43.3        &    39.8	     \\
$ ^{4}t_{0} $ &    108.3     &    108.5       & 	  58.0	      &	  58.5        &    53.4	     \\
$ ^{5}t_{0} $ &    135.1     &    135.8       &    72.4        &   72.5        &    66.5	     \\
$ ^{6}t_{0} $ &    161.4     &    161.8       &    86.5	      &	  87.1        &    79.5	     \\
$ ^{7}t_{0} $ &    187.6     &    188.1       &    100.5       &   100.5       &    92.4	     \\
\\
$ ^{2}t_{1} $ &	 1388.9     &    1402.6      &	  891.4       &   893.2       & 822.4		\\
$ ^{2}t_{2} $ &	 1855.4     &    1863.2      &	 1424.1	      &  1437.3       &  1348.2  	\\
$ ^{2}t_{3} $ &	 2435.5     &    2488.4      &   1710.2	      &  1718.2       &  1659.7 	\\
\hline 
\end{tabular}
\end{center}
\end{table}

\begin{table}
\begin{center}
\caption{Frequencies of the torsional modes $^l t_n$ in Hz for a
  neutron star model with entrainment, for three different phase
  transition densities $\rho_{ph}=1,5,10 \times 10^{13} \textrm{g
    cm}^{-3}$.  These results are obtained with the FD approach.
\label{tab2} }
\begin{tabular}{  c  c c c   }
\hline
                &           & $\rho_{ph}/ 10^{13}  \textrm{g cm}^{-3} $ &     \\
Mode            &   10      &  5                                     & 1   \\
\hline 
$ ^{2}t_{0}   $  &   25.0    &  20.8     &   16.5 \\
$ ^{3}t_{0}   $  &   39.0    &  32.8     &   26.1  \\
$ ^{4}t_{0}   $  &   53.1    &  44.0     &   34.9  \\
$ ^{5}t_{0}   $  &   66.3    &  54.9     &   43.6  \\
$ ^{6}t_{0}   $  &   79.2    &  65.6     &   52.1  \\
$ ^{7}t_{0}   $  &   92.0    &  76.2     &   60.5 \\ 
\\
$ ^{2}t_{1}   $  &   891.1   &  857.6    &  690.7  \\
$ ^{2}t_{2}   $  &   1422.7  &  1319.3   &  1096.0 \\
$ ^{2}t_{3}   $  &   1708.2  &  1936.6   &  1452.6 \\
\hline
\end{tabular}
\end{center}
\end{table}

\subsection{Crustal modes} \label{sec:cm}

For testing purposes, we study the torsional shear modes of a
non-magnetised star with two methods, the time evolution code and
solving a 1D eigenvalue problem (see Appendix~\ref{sec:cmAp}). Our
results are reported in Table \ref{tab1}, where we compare the mode
frequencies obtained with both our Time Domain (TD) and Frequency
Domain (FD) approaches.  The results obtained with these two methods
agree to better than one percent.  As expected, the effect of the high
neutron effective mass resulting from entrainment is important.  The
mode frequency is roughly decreased by a 50 percent with respect to
models without entrainment and tend to be closer to the results of
non-superfluid stars. In fact, the relative difference of models
including entrainment with respect 1-fluid models is about a 10
percent, in agreement with previous works \citep{2009CQGra..26o5016S,
  2012MNRAS.419..638P, 2013MNRAS.428L..21S}.
All the results reported in Table~\ref{tab1} follow the scaling
expected for the fundamental modes with respect to the harmonic index
$l$~\citep{2007MNRAS.374..256S}:
\begin{equation}
\nu \approx \sqrt{\left( l-1 \right) \left( l+2 \right)} \, . \label{eq:sc}
\end{equation}

The effect of the existence of a nuclear pasta layer is shown in Table
\ref{tab2}.  The mode frequencies decrease with $\rho_{ph}$ (the wider
pasta phase region, the lower frequency), as a consequence of the
average smaller shear modulus at the bottom of the inner crust (see
Fig.\ref{fig:mu}).  This trend is in agreement with the results
presented in \citet{2011MNRAS.417L..70S} and
\citet{2011MNRAS.418.2343G}.

We use a frequency domain code to determine also the eigenfunctions of
the crustal modes.  They will be used as initial conditions for the
time evolutions of magneto-elastic oscillations.  More precisely, we
consider the eigenfunctions of the fundamental modes, up to $l=20$,
and the first six overtones with $l=2$.

\subsection{Magneto-elastic waves}

\begin{figure*}
\begin{center}
\includegraphics[height=75mm]{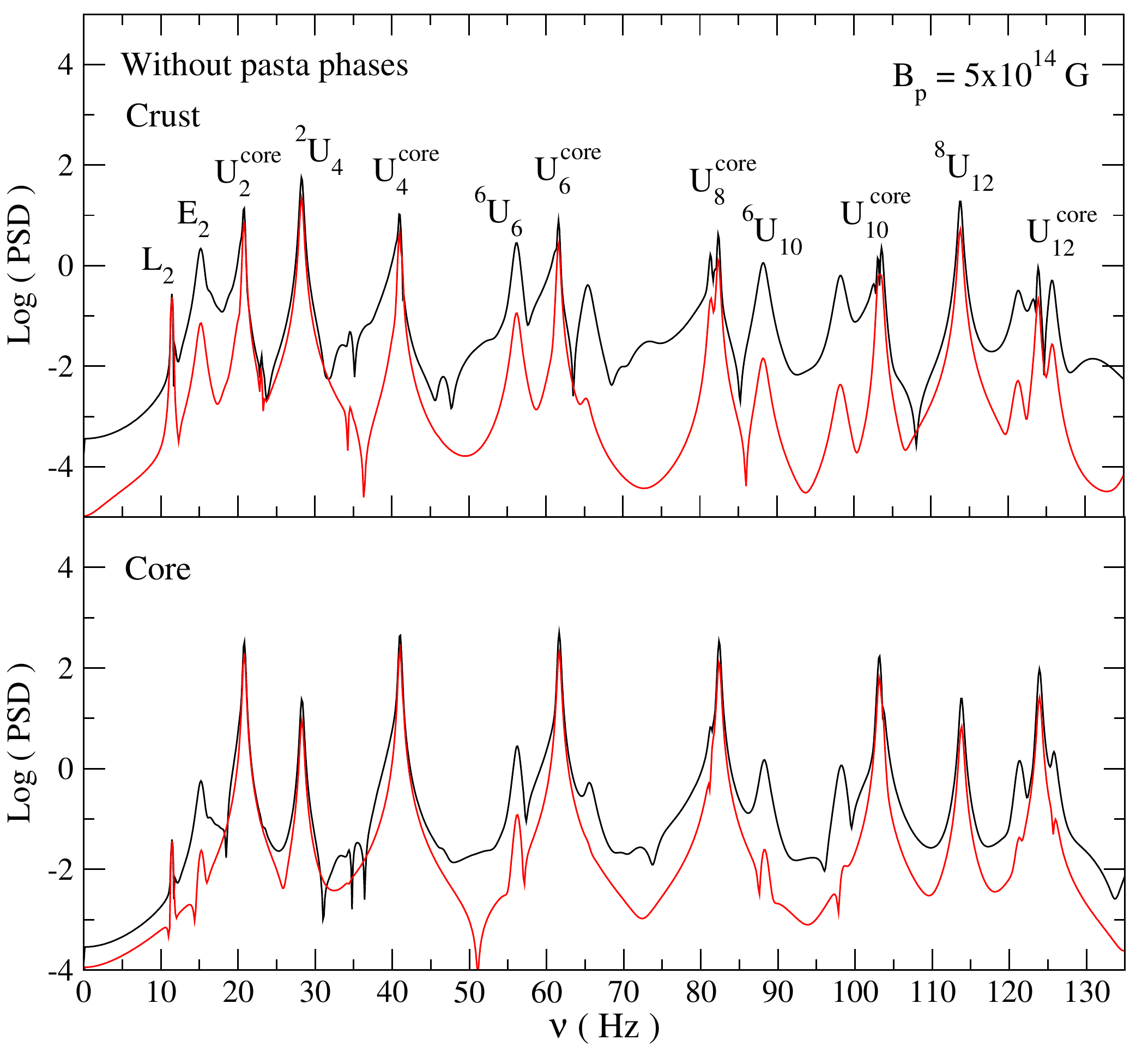} 
\includegraphics[height=75mm]{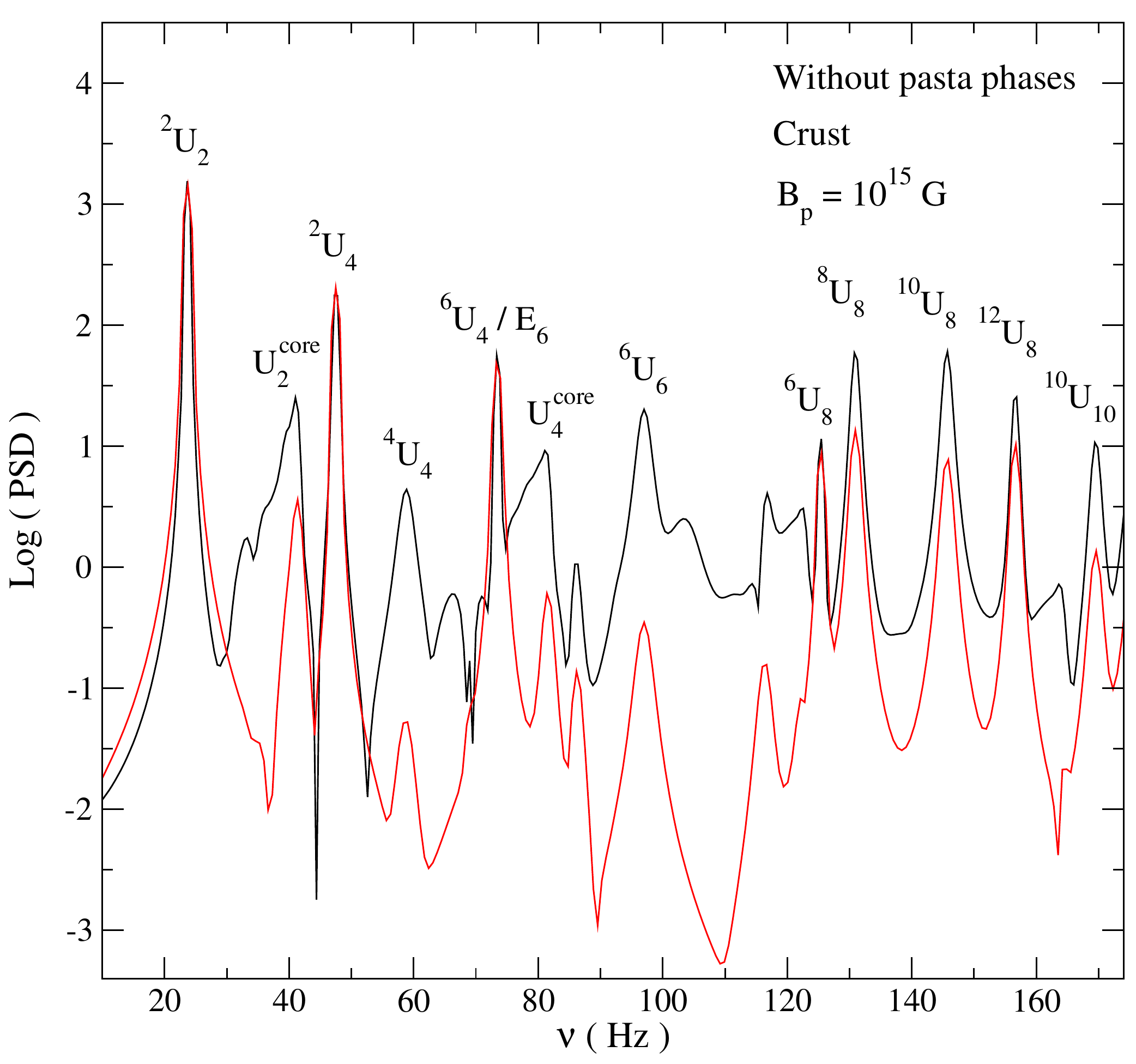} 
\caption{This figure displays the Power Spectrum Density (PSD) of a
  star without nuclear pasta and with two different magnetic field
  strength.  The left-hand panel shows an FFT taken for a star with
  $B_p=5\times10^{14}$G in two positions near the magnetic axis, in
  the crust (upper left-hand panel) and the core (lower left-hand
  panel).  The black line refers to an FFT performed for a 8~s time
  simulation, while the red line corresponds to the interval $
  1~\textrm{s} < t < 8~\textrm{s}$. For a model with $B_p= 10^{15}$G,
  the right-hand panel shows the results of an FFT performed in the
  crust near the magnetic axis, for the time interval $ 0~\textrm{s} <
  t < 2~\textrm{s}$ (black line) and $ 0.6~\textrm{s} < t <
  2~\textrm{s}$ (red line). The horizontal axis provides the mode
  frequency in Hz, while the vertical axis the logarithm of the PSD.
\label{fig:FFTP0}}
\end{center}
\end{figure*}

The interpretation of the low frequency QPOs as crustal modes
drastically depends on the magnetic field strength and the star
model. In strongly magnetised stars, crustal modes can live longer if
they appear in continuum gaps.  This situation is more feasible at low
frequencies where the continuum gaps are more likely present
\citep{2011MNRAS.410.1036V,2012MNRAS.420.3035V}.  As seen in
Sec. \ref{sec:cm}, a way to have a more dense population of shear
modes at lower frequencies is to assume an extended nuclear pasta
region (low shear) in the inner crust, in the density range $ 10^{13}
\textrm{g cm}^{-3} \lesssim \rho \lesssim 10^{14} \textrm{g cm}^{-3}$
\citep[see e.g.][]{2013PhRvC..88f5807S}.  Therefore, it is worth investigating
whether these modes survive in the spectrum of strongly magnetised
stars.

We first study the magneto-elastic oscillations in a neutron star
model without nuclear pasta.  As described in Sec. \ref{sec:cm}, we
initially excite several purely crustal modes and let the system
evolve. In this work, we consider a poloidal magnetic field which
strength varies in the range $10^{14}$G~$ \le B_p \le 2 \times
10^{15}$G.  To study the properties of the oscillations, we extract
the mode frequencies by using the FFT of the time evolved
perturbation.  We find that for $B_p \lesssim 4 \times 10^{14}$G, the
oscillation modes show the same characteristics of non-superfluid
models, which have been thoroughly discussed in the literature.  After
a transition, where many oscillations of the magnetic field continuum
are excited, the spectrum is characterised by the edge modes
(vibrations of the last open field lines), the upper modes (mostly
localised near the magnetic axis), and the oscillations confined into
the closed field line region.  For a stronger magnetic field, $B_p
\gtrsim 5 \times 10^{14}$G, there are new features in the spectrum
with some of them clearly showing a constant phase character.  This
kind of magneto-elastic waves have been already found by
\citet{2013PhRvL.111u1102G} and \citet{2014MNRAS.438..156P} and
recently studied more extensively by \citet{2016arXiv160507638G}.  One
interesting property of these modes is that they are not confined into
the core and their amplitude patterns reproduce, in many cases, the
longitudinal nodal lines of pure torsional crustal oscillations.

In the literature, there is a proliferation of notation for the mode
classification \citep[see Table 1 in][]{2016arXiv160507638G}.  In this
work, we prefer to converge toward the notation recently introduced by
\citet{2016arXiv160507638G} in order to make easier future
comparisons. We denote by E$_n$ and U$_n$, respectively, the edges and
the upper turning points of the continuum bands. The oscillations of
the closed magnetic field lines will be labelled as
L$_{n}$ \footnote{Note that in \citet{2014MNRAS.438..156P} the edge
  modes of the continuum were called L$_n$ instead of E$_n$, and the
  closed field line modes C$_n$ instead of L$_n$.}.  In addition, we
add the superscript `core' to specify the modes which remain mostly
confined into the core, while the label $n$ now corresponds to the
number of maxima that a mode has along the field lines.  For example,
in the new notation, an U$_2 ^{\rm core}$ mode corresponds to the
antisymmetric U$_0 ^{(-)}$ mode of \citet{2014MNRAS.438..156P}. We
label the global magneto-elastic oscillations by $^l$U$_{n}$, where
$n$ denotes, as for the core confined modes, the number of maxima
present along the magnetic field lines, while the index $l$ describes
the angular structure of the mode pattern. Therefore, a
magneto-elastic mode with $l=2$ has the same angular structure of a
purely $l=2$ crustal mode\footnote{Note that in
  \citet{2014MNRAS.438..156P} an $^l$U$_{n}$ mode is called $^l
  t_{n}^{\ast}$}.

\begin{figure}
\begin{center}
\includegraphics[trim = 28mm 12mm 34mm 12mm, clip, height=39mm]{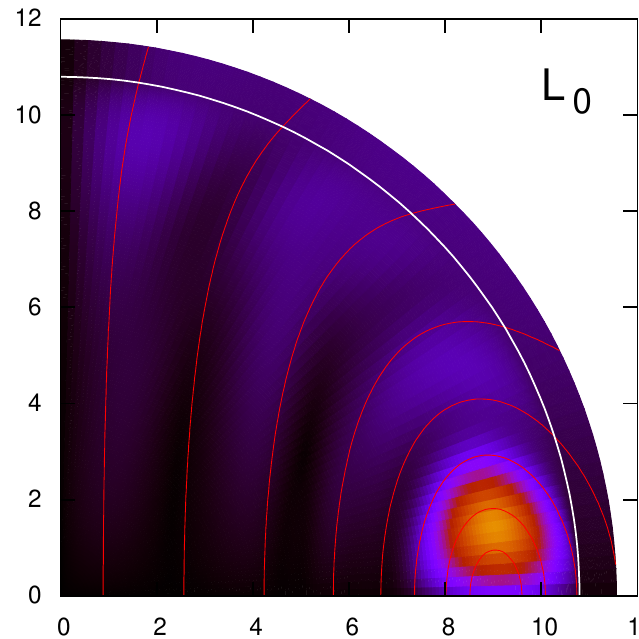}  
\includegraphics[trim = 28mm 12mm 34mm 12mm, clip, height=39mm]{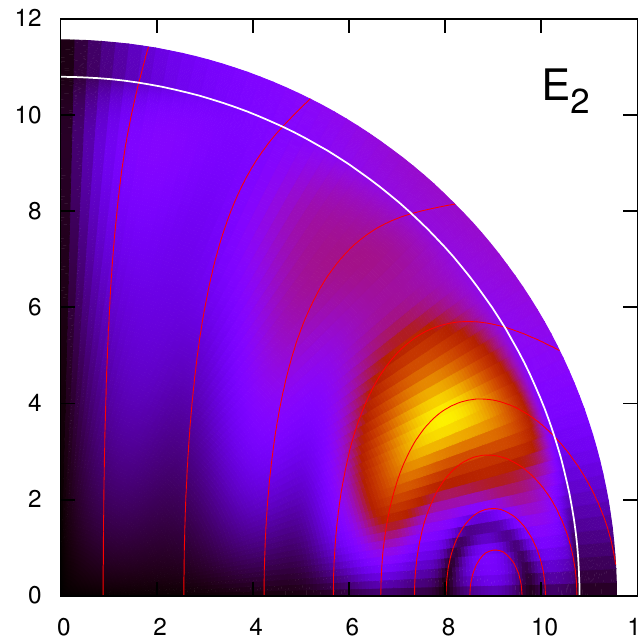}   \\
\includegraphics[trim = 28mm 12mm 34mm 12mm, clip, height=39mm]{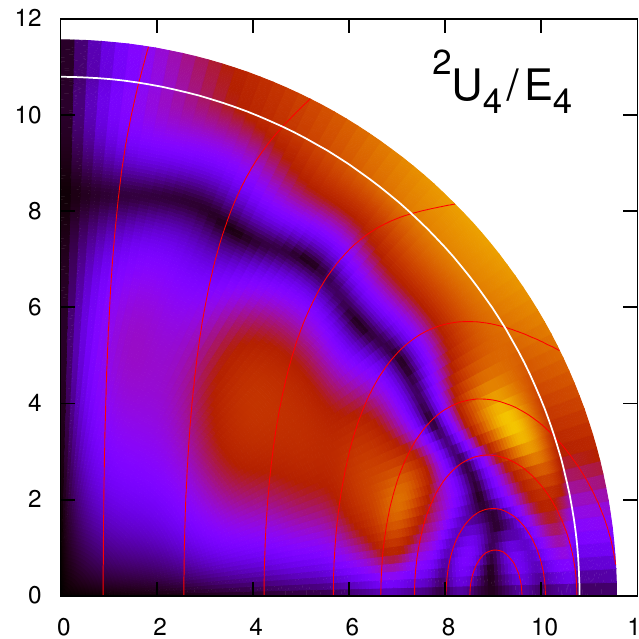}  
\includegraphics[trim = 28mm 12mm 34mm 12mm, clip, height=39mm]{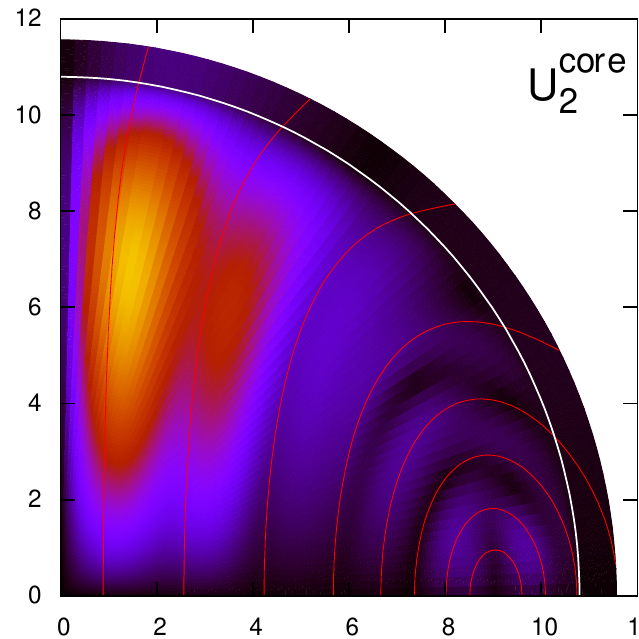}   \\
\includegraphics[trim = 28mm 12mm 34mm 12mm, clip, height=39mm]{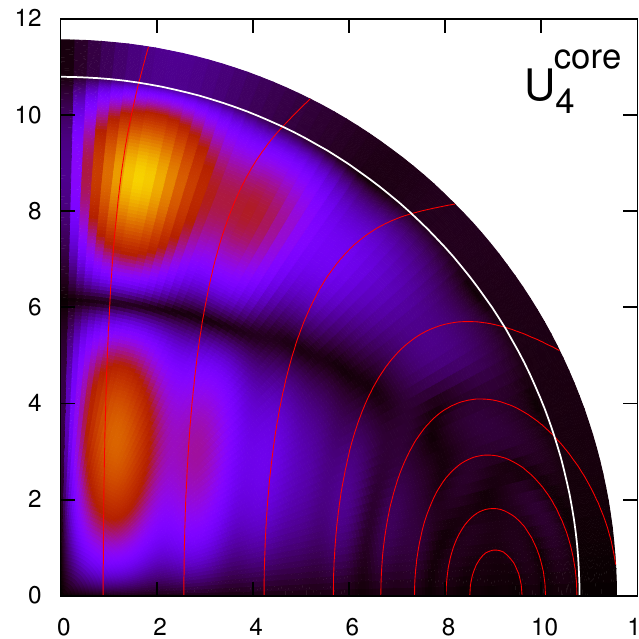}  
\includegraphics[trim = 28mm 12mm 34mm 12mm, clip, height=39mm]{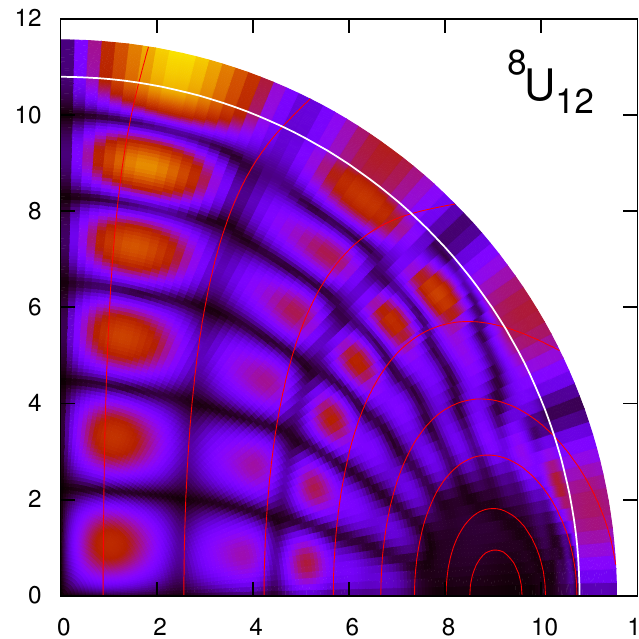} 
\caption{Effective 2D-FFT of six magneto-elastic waves for a model
  with $B_{p}=5 \times 10^{14}$G and without nuclear pasta.  We show
  the absolute value of the amplitude of $\xi^{\phi}$ as determined in
  the orthonormal basis.  From the spectral features of
  Fig.~\ref{fig:FFTP0} we select the following modes L$_{0}$, E$_{2}$,
  $^{2}$U$_{4}$/E$_{4}$, U$_{2}^{\rm core}$, U$_{4}^{\rm core}$,
  $^8$U$_{12}$. The magnetic field lines are shown in red, while the
  crust/core interface with a white line.
\label{figA}}
\end{center}
\end{figure}

\begin{figure}
\begin{center}
\includegraphics[trim = 28mm 12mm 34mm 12mm, clip, height=39mm]{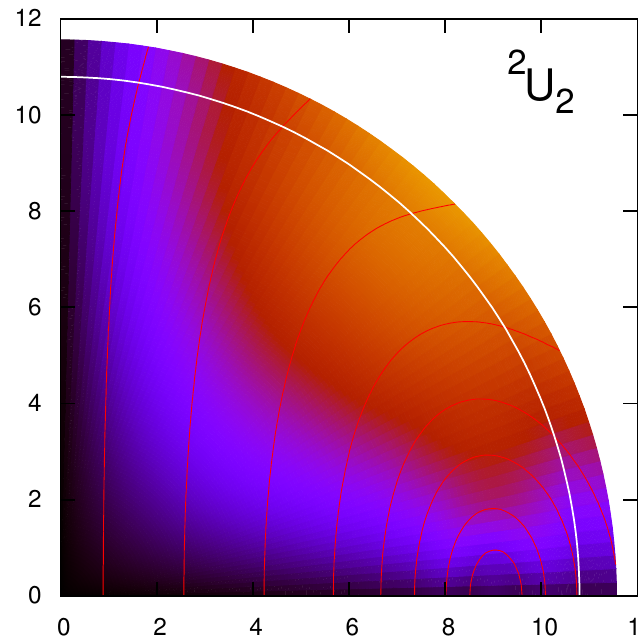}  
\includegraphics[trim = 28mm 12mm 34mm 12mm, clip, height=39mm]{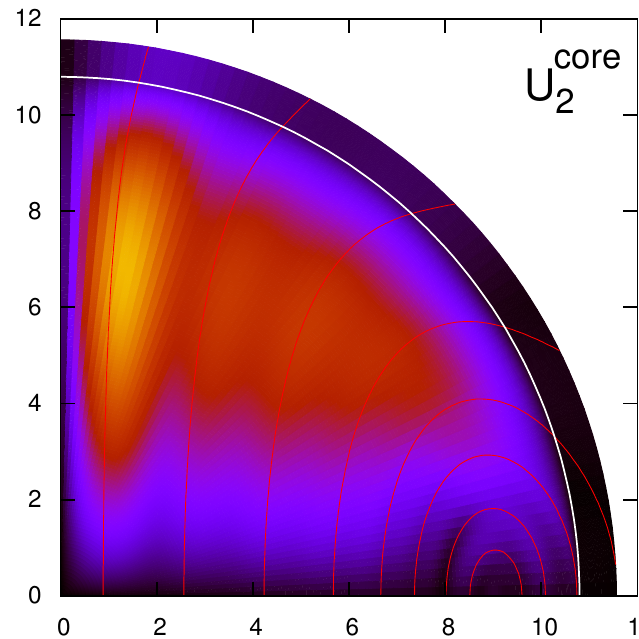}  \\
\includegraphics[trim = 28mm 12mm 34mm 12mm, clip, height=39mm]{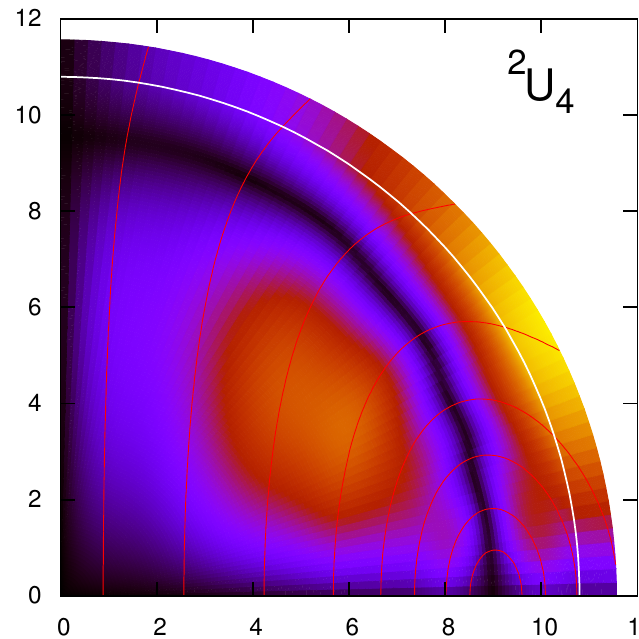}  
\includegraphics[trim = 28mm 12mm 34mm 12mm, clip, height=39mm]{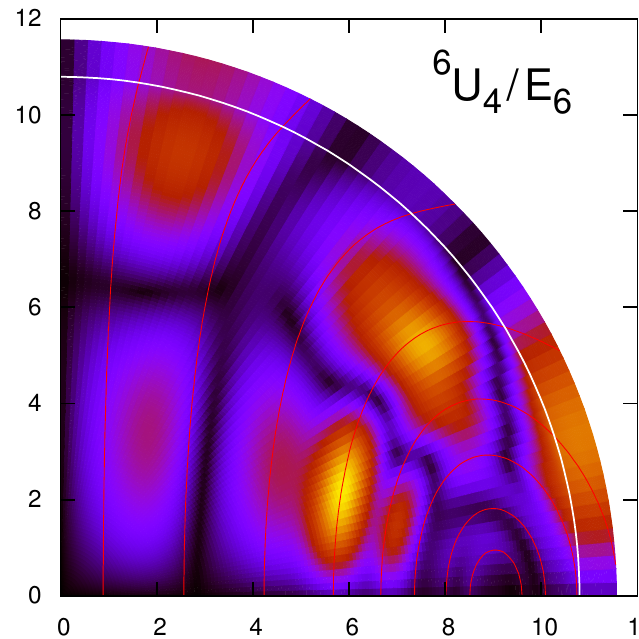}   \\
\includegraphics[trim = 28mm 12mm 34mm 12mm, clip, height=39mm]{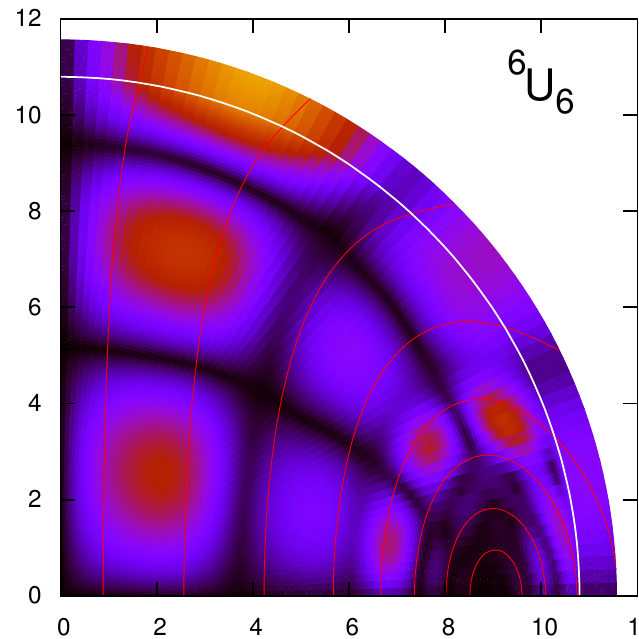} 
\includegraphics[trim = 28mm 12mm 34mm 12mm, clip, height=39mm]{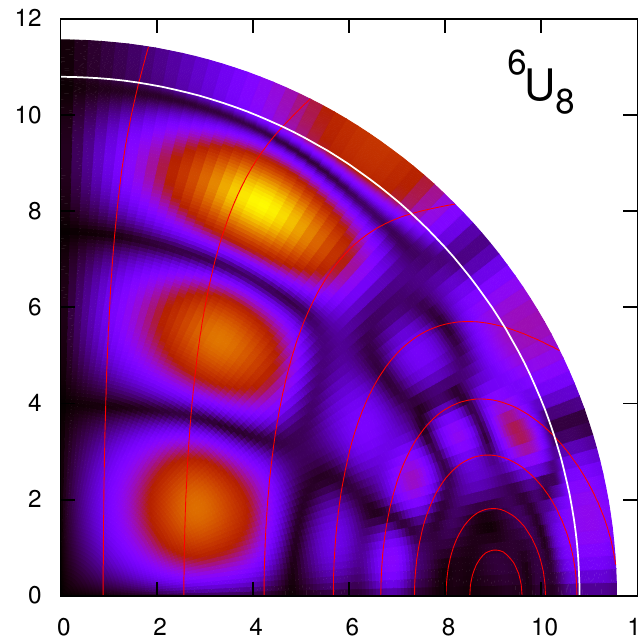}  
\caption{Same as Fig. \ref{figA} for $B_{p}=10^{15}$G.  The modes
  selected in this case are $^2$U$_{2}$, U$_{2}^{\rm core}$,
  $^2$U$_{4}$ , $^6$U$_{4}$ /E$_6$, $^6$U$_{6}$ and $^6$U$_{8}$.  
\label{figA15}}
\end{center}
\end{figure}
 
For a star with $B_{p}=5 \times 10^{14}$G, we show in
Fig.~\ref{fig:FFTP0} an FFT of the Lagrangian displacement taken in
two positions near the magnetic axis, in the crust (left-upper panel)
and in the core (left-lower panel).  To identify the long living
oscillations, we have determined the FFT for two different time
intervals: the entire evolution $t < 8$~s and for $ 1~\textrm{s} < t <
8$~s.  As pointed out before, many modes are initially excited, but
some of them are already significantly damped after one second. The
core confined U$^{\rm core}$ mode, the L modes, and a couple of global
magneto-elastic modes ($^2$U$_{4}$ and $^8$U$_{12}$) persist longer
and appear to be only weakly damped by numerical dissipation. In
particular, the Upper modes are clearly dominant in the core.  The 2D
patterns of six selected magnetic-elastic modes are shown in
Fig. \ref{figA}. As expected, the E and U$^{core}$ modes are mainly
confined into the core, due to the velocity step present at the
crust/core interface, which results in a low efficiency in the
transmission of Alfven waves.  However, already at this magnetic field
strength, the magneto-elastic modes $^2$U$_{4}$ and $^8$U$_{12}$ have
a small damping, as seen in Fig.~\ref{fig:FFTP0}, and clearly reach
the surface. In particular, the 2D-pattern of the $^2$U$_{4}$
oscillation shows also the presence of an E$_{4}$ mode. Since these
two oscillations have similar frequency, the routine that extracts the
pattern amplitude from the time simulation is not able to discern
between the two modes. Also for the E$_{2}$ mode, the 2D pattern
indicates the presence of a global oscillation with amplitude much
smaller than the edge mode. For stronger magnetic fields, this global
oscillation will emerge as an $²$U$_2$ magneto-elastic wave.

\begin{figure}
\begin{center}
\includegraphics[trim = 28mm 12mm 24mm 12mm, clip, height=36mm]{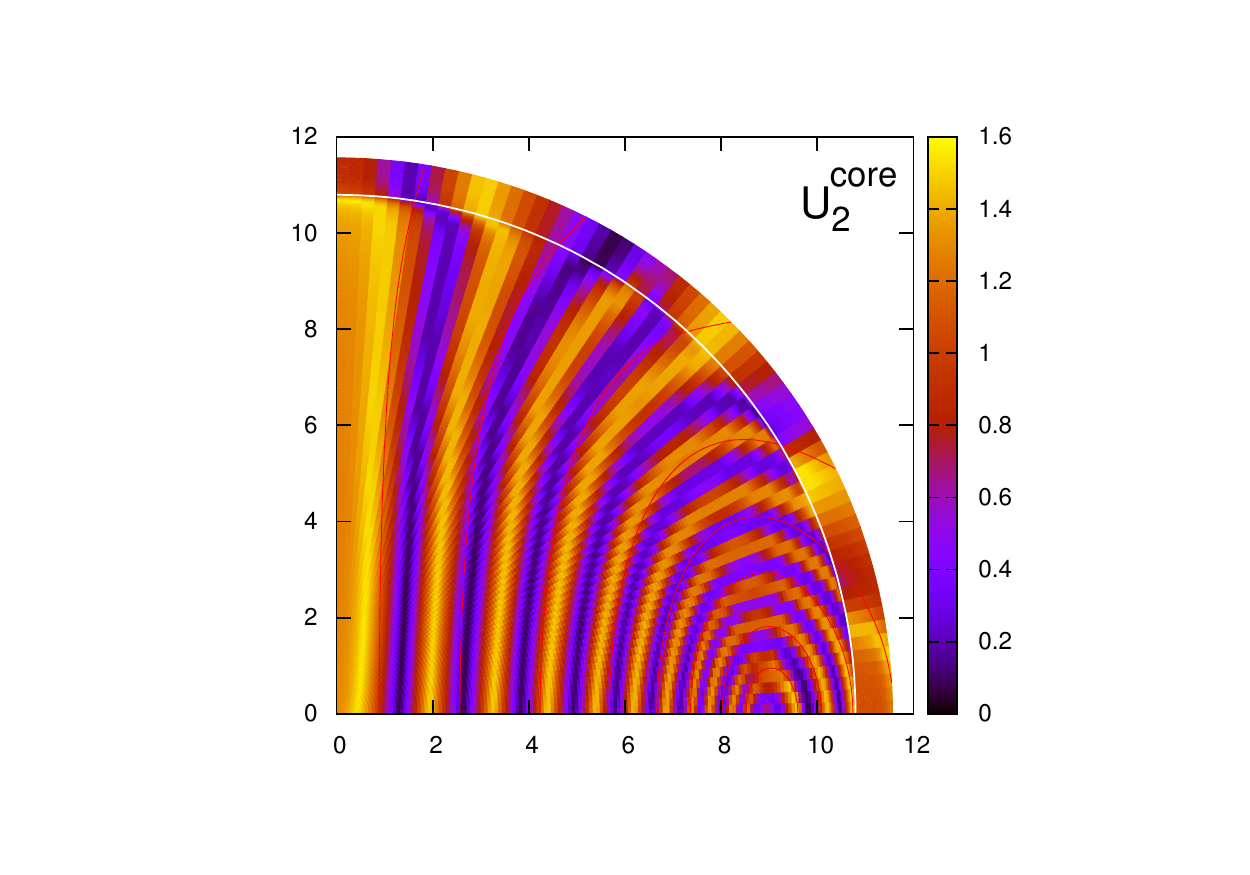}  
\includegraphics[trim = 28mm 12mm 24mm 12mm, clip, height=36mm]{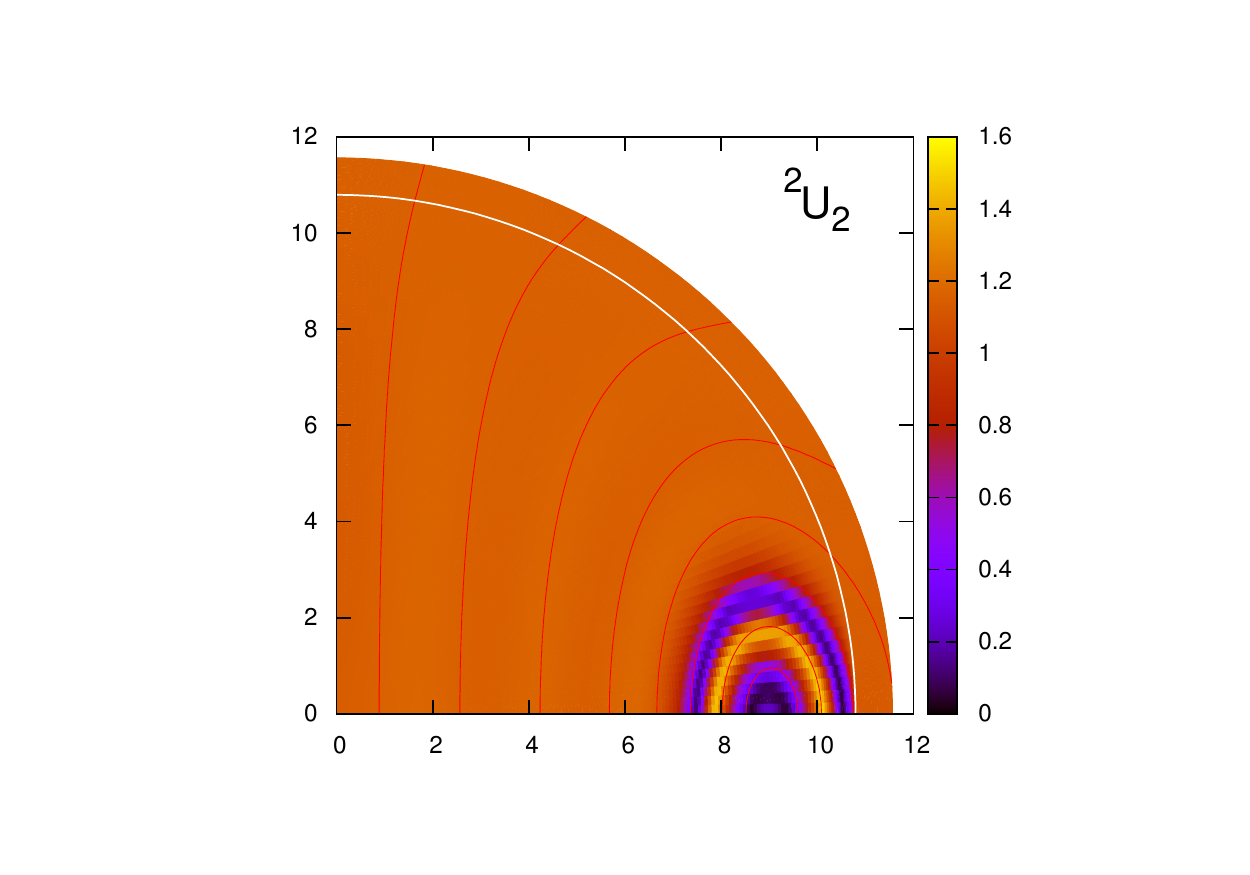}   \\
\includegraphics[trim = 28mm 12mm 24mm 12mm, clip, height=36mm]{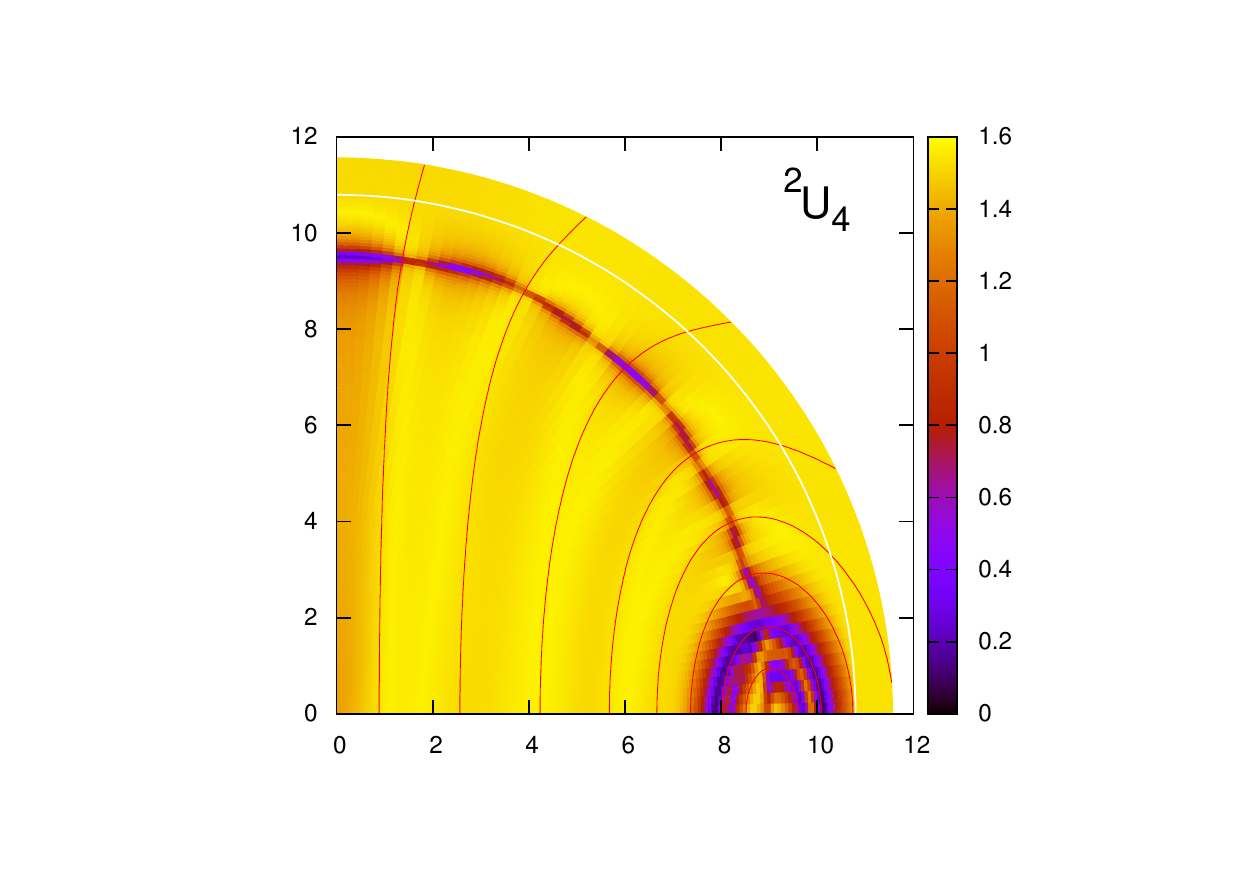}  
\includegraphics[trim = 28mm 12mm 24mm 12mm, clip, height=36mm]{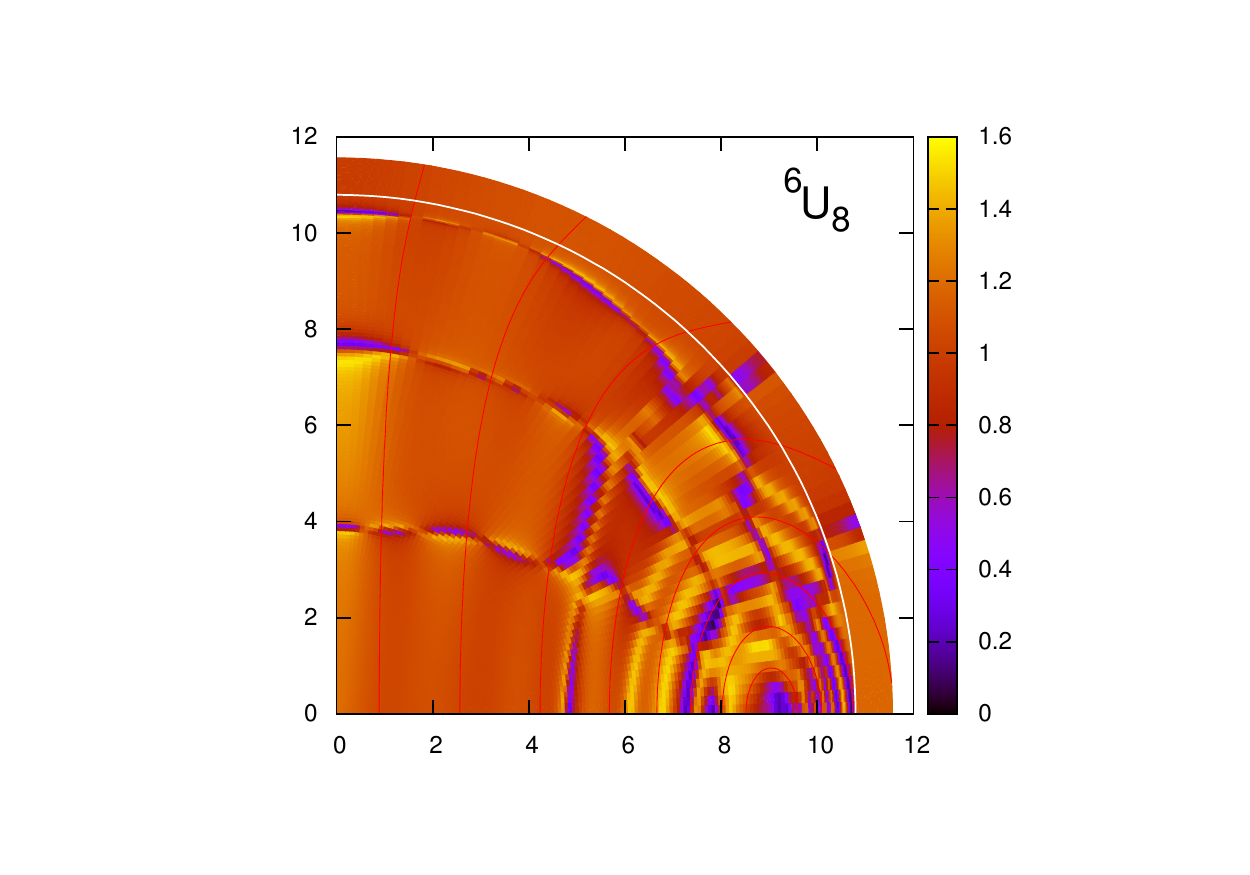}  
\caption{For some of the modes shown in Fig. \ref{figA15}, this figure
  displays the absolute value of the phase, given in radiants. The
  star is a model without nuclear pasta and $B_p =10^{15}$G. The modes
  represented are the core confined U$_{2}^{\rm core}$ mode and the
  following three global magneto-elastic oscillations $^2$U$_{2}$,
  $^2$U$_{4}$ and $^6$U$_{8}$.
\label{fig:phase}}
\end{center}
\end{figure}

By gradually increasing $B_p$, we find that the core confined U modes
are less excited while the global magneto-elastic oscillations become
dominant.  Tracking these modes with varying magnetic field strength,
we find that the long-living global magneto-elastic oscillations are
actually at the frequency expected for the E$_n$ modes.  In our
  stellar model these two kinds of oscillation modes have similar
  frequencies. The E$_n$ modes are more excited when the magnetic
  field is weaker, while the global magneto-elastic oscillations
  dominate for stronger magnetic fields.

 For a star with $B_p = 10^{15}$G, we show in
Fig.~\ref{fig:FFTP0} an FFT taken close to the magnetic axis and in
the crust for two time intervals, $ 0~\textrm{s} < t < 2$~s (black
line) and $0.6~\textrm{s} < t < 2$~s (red line), respectively.
Several magneto-elastic modes are now excited with different damping
times. The 2D patterns (Fig.~\ref{figA15}) show that the
magneto-elastic oscillations reach the surface and have more angular
structure than the core confined U modes. Note that in the
$^6$U$_{4}$, $^6$U$_{6}$ and $^6$U$_{8}$ oscillations, the difference
between the various local maxima of the amplitude is at most a factor
of five.
  
\begin{figure}
\includegraphics[height=75mm]{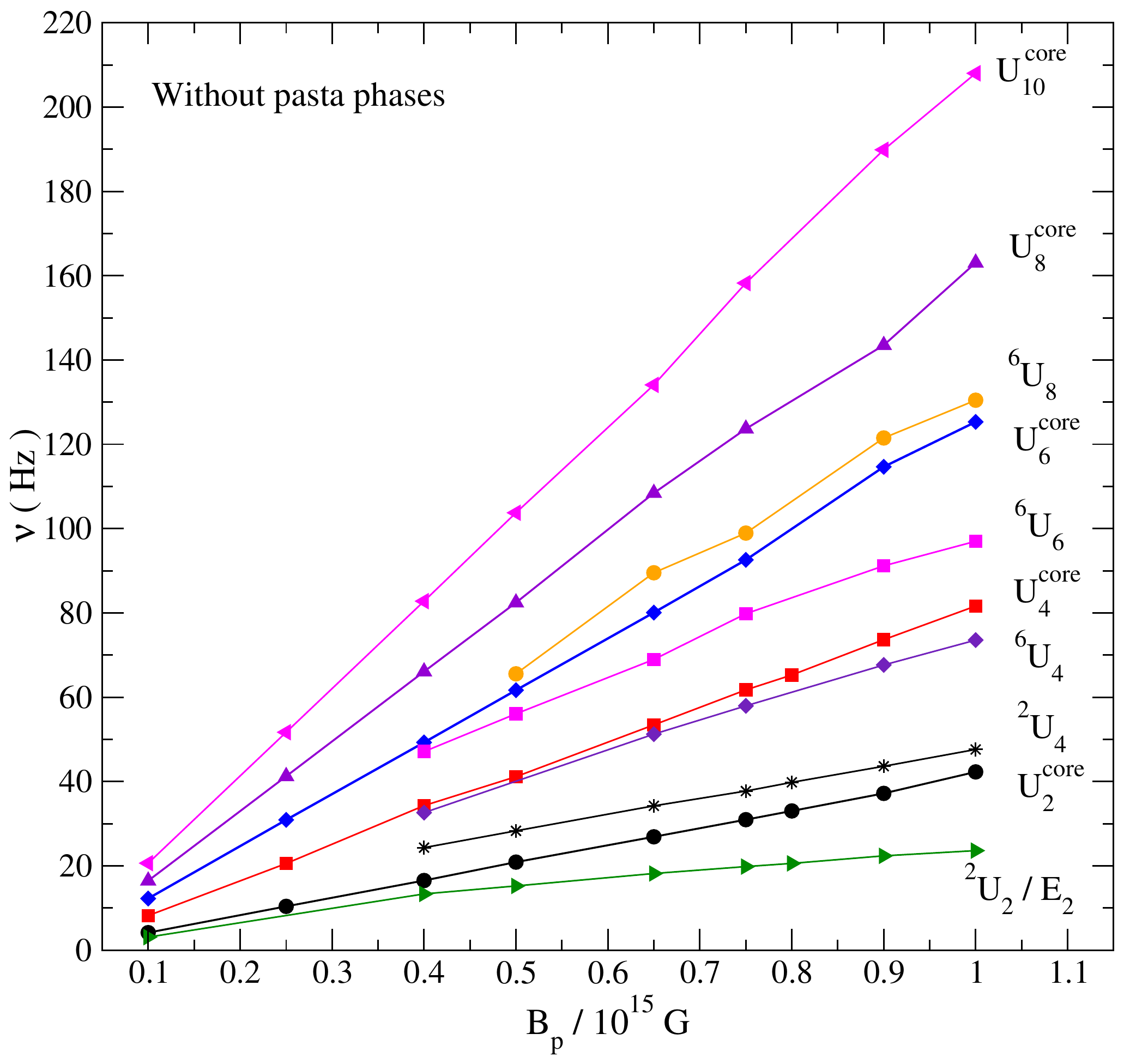}    
\caption{Antisymmetric magneto-elastic modes for a model without
  nuclear pasta.  The figure shows the frequency dependence of several
  core-confined U-modes and global magneto-elastic oscillations on the
  magnetic field $B_p$ (see legend).
\label{fig:PO-freq}}
\end{figure}
As mentioned above, a way to identify a global oscillation which does
not arise from the continuum is to study the oscillation phase
\citep{2013PhRvL.111u1102G, 2016arXiv160507638G}. If it is constant,
each fluid element oscillates in phase indicating a global coherent
character. If instead a mode belongs to the continuum, the fluid
element associated with a magnetic field line must oscillate out of
phase with the neighbour lines. Calculating the phase of various
magneto-elastic oscillations we find some modes which clearly exhibit
a coherent character. However, it is not always easy to identify a
constant phase in the mode overtones, as other modes with similar
frequency can appear in the same pattern and numerical noise might be
present around the nodal lines.  We show in Fig. \ref{fig:phase} the
phase of some modes for a star without nuclear pasta and with $B_p =
10^{15}$G.  The core confined U$_2^{\rm core}$ mode exhibits a highly
variable phase, while the other global magneto elastic modes, which
reach the surface, show a nearly constant phase. This is specially
evident for the $^2$U$_{2}$ mode, and less clear for the other two,
due to the existence of nodal zones where the amplitude of the
oscillation is zero and the color scale does not reflect correctly the
phase. It is also seen how the equatorial ring confined into the
closed field lines is effectively decoupled from the rest of the star.

Finally, we have checked that the frequency of the magneto-elastic
modes scales linearly with the magnetic field strength. By tracking
the mode frequency and identifying its 2D-pattern for models with
different magnetic field strengths, we find the results shown in
Fig. \ref{fig:PO-freq}.  In the explored range, the mode frequencies
have a clear linear trend with respect to $B_p$. As found also by
\citet{2016arXiv160507638G} the frequencies of the global
magneto-elastic oscillations cannot be determined for $B_p <
4\times10^{14}$~G, due to numerical limitation.  Therefore, we cannot
follow their behaviour for weaker magnetic fields.

To compare our results with the literature, we must determine the
relation between the magnetic field at the pole $B_p$ and the quantity
introduced by \citet{2016arXiv160507638G}. The authors of this work
provide the results as a function of the magnetic field strength,
$\bar B $, of a uniformly magnetised sphere, which has the same
magnetic dipole moment as the stellar model under exam. Furthermore,
to rescale the value to a standard model with a radius of 10~km, they
used the following equation (Eq. 19 in \citet{2016arXiv160507638G})
 \begin{equation}
 \bar B = \frac{ m_{dip}}{ \left( 10~\textrm{km} \right)^3} \left( \frac{10~\textrm{km}}{R} \right)^2 \, , 
 \label{eq:Bhat}
 \end{equation}
where $m_{dip}$ is the magnetic dipole moment of the background
model. For the star used in our work Eq.~(\ref{eq:Bhat}) provides
$\bar B = 0.425 B_p$. For equivalent $\bar B$, we find that our
results are consistent with those of \citet{2016arXiv160507638G}.  The
main characteristics of the oscillation spectrum and the 2D amplitude
patterns are very similar. However, we cannot directly compare the
oscillation frequencies, because the stellar models used in these two
works are different.

\subsubsection{Magnetar oscillations with a nuclear pasta layer}

 Now we discuss the possibility of a region with nuclear pasta phase.
 In Sec. \ref{sec:back} we presented our approximated model for the
 shear modulus, which smoothy decreases toward the crust/core
 interface in order to mimic the reduced elasticity expected in the
 non-uniform pasta structures.  We explored a wide range of the pasta
 phase transition density, $10^{13} \textrm{g cm}^{-3} \leq \rho_{ph}
 \leq 10^{14} \textrm{g cm}^{-3}$.  In general, we find that the
 excitation of the continuum is more evident during the initial
 transition period for these models than for models without nuclear
 pasta, but the system quickly re-distributes the initial energy among
 the various modes.  Depending on the magnetic field strength, the
 long living modes are the modes at the turning points and edges of
 the continuum for $B_p \lesssim 5\times 10^{14}$G, and the global
 magneto-elastic modes for stronger magnetic fields, $B_p > 5 \times
 10^{15}$G.
\begin{table}
\begin{center}
\caption{\label{tab:ti} Frequencies of the antisymmetric
  magneto-elastic oscillations (in Hz) for a neutron star with
  entrainment and magnetic field $B_p=10^{15}$G. The second column
  refers to the model without nuclear pasta, while the last three
  columns present the results for different pasta phase transition
  densities, $\rho_{ph}= 1,5, 10 \times 10^{13} \textrm{g
    cm}^{-3}$. These results are determined from the FFT of the time
  evolution results.
 \label{tab4} }
\begin{tabular}{  c  c  c c  c c  c }
\hline
\hline
                   &          &    & $\rho_{ph}/ 10^{13}  \textrm{g cm}^{-3} $ &  \\
           Mode               &   No pasta    &  10      &  5   &  1     \\
\hline 
 $^2$U$_{2} $       &  23.8   &    23.3  &   22.8   &  22.1       \\   
 $^2$U$_{4} $       &  47.4   &    46.1  &   47.7   &  47.4       \\ 
 $^6$U$_{4} $       &  73.6   &    72.5  &   72.8   &  71.9       \\
 $^6$U$_{6} $       &  97.1   &  100.3   &   93.7   &  97.3       \\
 $^6$U$_{8} $       & 125.6   &  121.6   &  119.4   &  120.3      \\
 $^{10}$U$_{8}  $    & 145.7   &  137.8   &  146.8   &  144.4      \\
 $^{10}$U$_{10}  $   & 169.4   &  170.4   &  171.6   &  168.6      \\
\hline
\end{tabular}
\end{center}
\end{table}

\begin{figure*}
\begin{center}
\includegraphics[height=75mm]{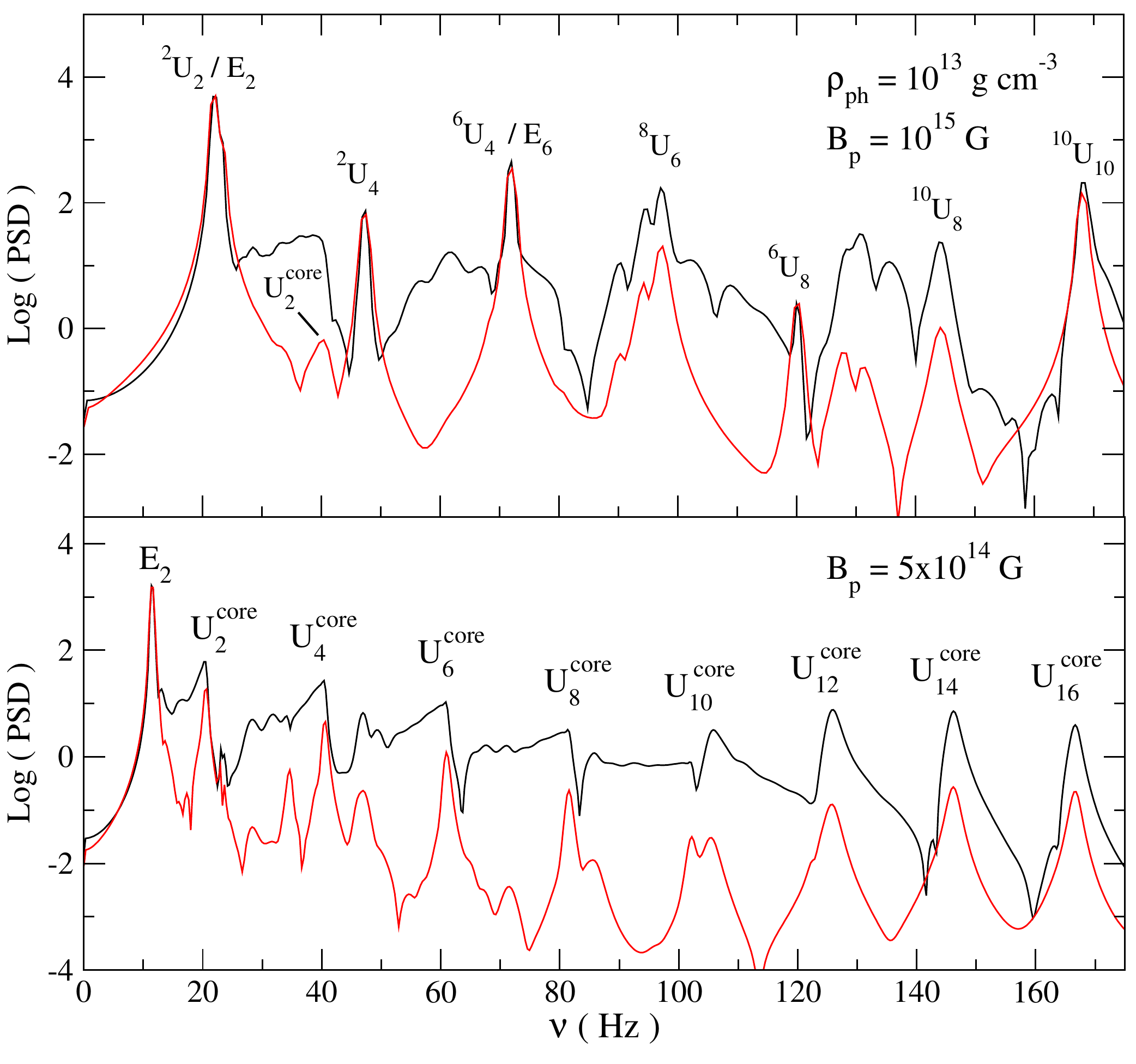} 
\includegraphics[height=75mm]{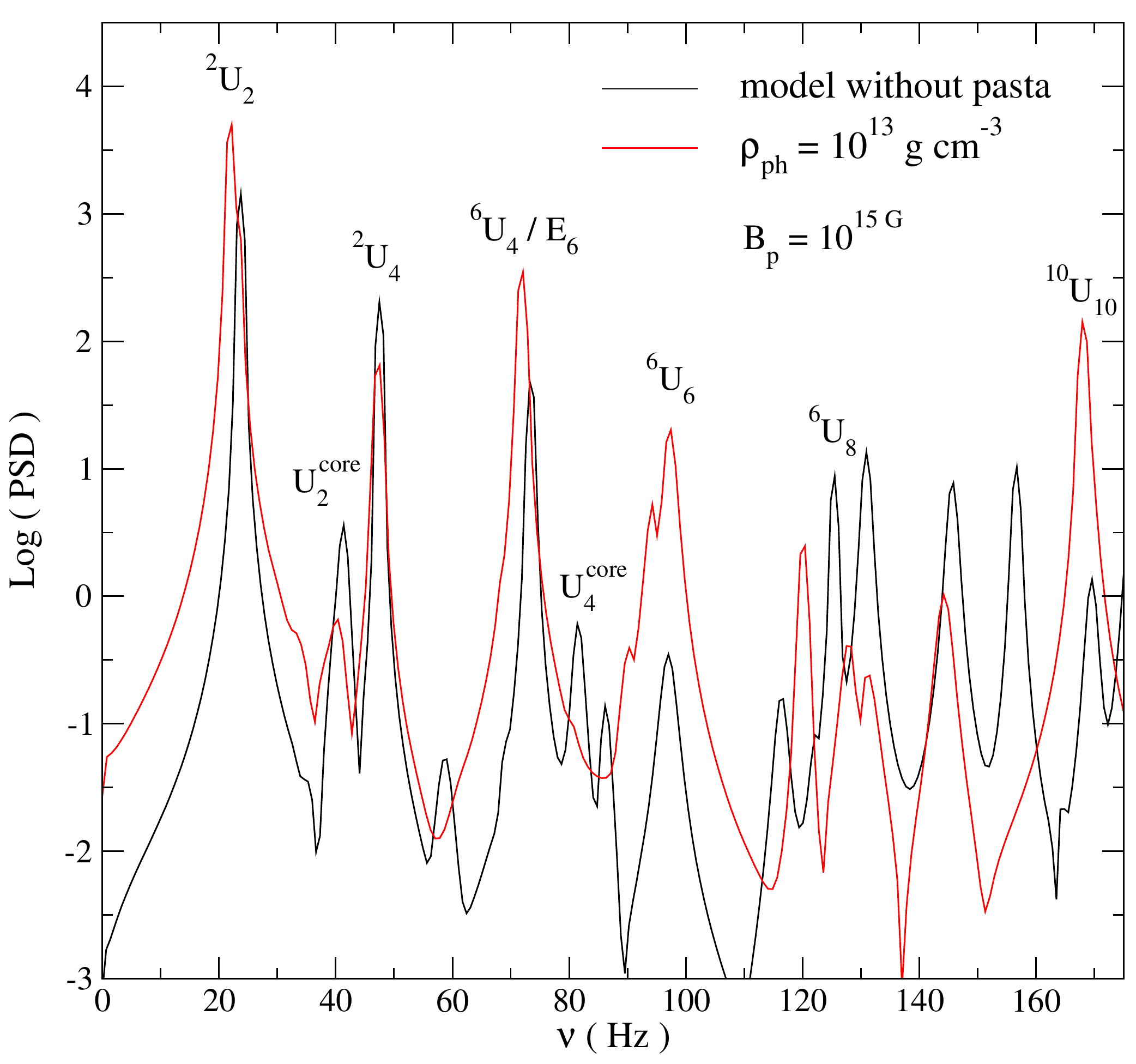} 
\caption{ Power spectrum density of a star's model with nuclear pasta
  transition $\rho_{ph} = 10^{13}~\textrm{g cm}^{-3}$.  The FFTs are
  taken in the crust near the magnetic axis.  The left-hand panel
  displays the FFT for the two intervals: $0 ~\textrm{s} \leq t \leq
  2~\textrm{s}$ (black line) and $0.6 ~\textrm{s} \leq t \leq 2
  ~\textrm{s}$ (red line).  The star with $B_p=5\times10^{14}$G is
  shown on the lower-left panel, while the $B_p= 10^{15}$G case on the
  upper-left panel.  The right-hand panel compares the FFT of a model
  with $\rho_{ph} = 10^{13}~\textrm{g cm}^{-3}$ (black line) with a
  star without nucelar pasta (red line). Both models have $B_p=
  10^{15}$G and the FFT is taken for $0.6 ~\textrm{s} \leq t \leq
  2~\textrm{s}$.
\label{fig:FFTP13}}
\end{center}
\end{figure*}

We focus on a star with $\rho_{ph} = 10^{13} \textrm{g cm}^{-3}$,
which has the widest nuclear pasta layer within the models considered.
In the left panel of Fig.~\ref{fig:FFTP13}, we compare the FFT of the
Lagrangian displacement for $B_p = 5\times 10^{14}$G and $B_p =
10^{15}$G. The FFTs are taken in the crust near the magnetic axis and
in two different time intervals, for the entire evolution $t = 2 $~s
(black line) and for $ 0.6~\textrm{s} <t < 2$~s (red line).  The main
oscillations excited in the star with $B_p = 5\times 10^{14}$G are
clearly the E$_2$ and U$_n^{\rm core}$ modes. Compared with the model
without nuclear pasta, it seems that less magneto-elastic modes are
excited in this frequency range (compare Fig.~\ref{fig:FFTP0} and
Fig.~\ref{fig:FFTP13}).  For $B_p = 10^{15}$G (top panel), more
magneto-elastic modes show a persistent character or in general a
smaller damping than for the $B_p = 5\times 10^{14}$G case.  In the
right panel of Fig.~\ref{fig:FFTP13} we compare the FFTs of two
models, with and without pasta phase, with the same magnetic field,
$B_p = 10^{15}$G.  The frequency of each magneto-elastic oscillation
is quite similar, especially for modes with higher amplitude.  In the
range $ 120 \textrm{Hz} < \nu < 160$~Hz there is a richer spectrum in
the model without nuclear pasta, with modes showing more nodal lines
along the $\theta$ coordinate (higher harmonic index $l$), but
maintaining the same radial structure (see also Fig.~\ref{fig:FFTP0}).
It is likely that in some frequency range different initial conditions
can excite different overtones, but the frequency of the dominant
modes seems to be closer to models without nuclear pasta.

Considering various models with different $\rho_{ph}$ transition, we
find that the spectrum maintains the same basic structure of models
without nuclear pasta, and that the frequency of global
magneto-elastic modes barely changes with $\rho_{ph}$ (especially for
the first three magneto-elastic oscillations).  In Table \ref{tab4} we
provide the mode frequencies of the magneto-elastic modes for a model
with $B_p = 10^{15}$G and for different pasta phase transition
densities $\rho_{ph}$.  We do not show in this table the core confined
U modes, as they are less excited into the crust at this magnetic
field strength. In any case, their frequencies are very close to those
of models without nuclear pasta.

\section{Conclusions} \label{sec:conc}

The interpretation of magnetar QPOs as driven by the magneto-elastic
oscillations is a well established scenario, and it is therefore
important to investigate all the physical effects which can modify the
oscillation spectrum.  In this work we have studied the torsional
oscillations of relativistic superfluid magnetars, where the magnetic
field has a purely poloidal geometry.

The main properties of the spectrum for stars with and without nuclear
pasta are similar to the results obtained by
\citet{2013PhRvL.111u1102G} and \citet{2014MNRAS.438..156P}, and
consistent with \citet{2016arXiv160507638G}.  In the magnetic field
range explored in this work, we find two different families of
magneto-elastic modes: a first class of core confined oscillations and
a second family of modes which penetrate the crust and reach the
surface. The modes of the first class show a no-constant phase and can
be associated with the oscillations at the turning points and edges of
the continuum bands. The modes which reach the surface have instead a
more structured angular pattern which reminds that of the crustal
modes, and many of them show a constant phase.  We find that these two
classes coexist in our simulations, but the core confined modes are
dominant at `weak' magnetic fields, typically when $B_p \lesssim 4
\times 10^{14}$G, while the global magneto-elastic oscillations
dominate at stronger magnetic fields, roughly for $ B_p \gtrsim 5
\times 10^{14}$G.  Interestingly, if we linearly extrapolate the
frequencies of the edge modes for stronger magnetic fields, we find
that the magneto-elastic modes with higher amplitude, which persist
longer in the evolution, reside close to the edge mode
frequency. Imprints of edge modes are in fact found in the 2D
oscillation patterns.

Although the crustal modes densely populate, in models with nuclear
pasta and strong entrainment, the low frequency part of the spectrum
where the continuum gaps are more likely present, we do not find any
direct imprints of a fundamental crustal mode for stars with $B_p \ge
10^{14}$G.  The magneto-elastic oscillations that reach the star's
surface in fact depend linearly on the magnetic field strength. This
behaviour should exclude a direct identification of these
magneto-elastic modes with purely crustal modes, as the frequency
variation for the magneto-corrected crustal modes is very small for
the magnetic field strength considered in this
work~\citep{2007MNRAS.375..261S, 2012MNRAS.420.3035V}.  Furthermore,
the basic structure of the magneto-elastic spectrum does not seem to
depend strongly on the presence of nuclear pasta, despite the various
models have crustal modes with very different frequencies.

We also explored the high frequency QPOs, around 625 Hz, which can be
in the frequency range of the first overtone of torsional crustal
modes.  Our simulation suggests that the first crustal mode overtone
used to excite the time evolution is efficiently absorbed by the
continuum when the pasta phase is present and $B_p \gtrsim 4\times
10^{14}$G. Therefore the identification of this mode with the 625 Hz
QPOs appears more difficult in models with nuclear pasta. This can be
due to the smoother transition of the shear modulus across the
crust/core interface which facilitates the wave transmission.
However, we think that the identification of the high frequency QPOs
require a more complex analysis of the spectrum which involves both
polar (spheroidal) and axial (torsional) oscillations. This is a key
requisite for studying more realistic magnetic field configurations
with mixed poloidal-toroidal geometry. With the presence of a toroidal
magnetic field component, the polar and axial perturbations couple and
the spectrum can change significantly.  For instance, there are
indications that due to this coupling the continuum spectrum can
disappear \citep{2012MNRAS.423..811C}. This result is in part
expected. Works in tokamak and solar physics have in fact shown that
the coupling between Alfv\'{e}n, gravity and pressure waves can open
gaps in the continuum bands \citep[see for
  instance][]{2008PhPl...15e5501H, 2011A&A...532A..94B}

\section*{Acknowledgements}
A.P. acknowledges support from the European Union under the Marie
Sklodowska Curie Actions Individual Fellowship, grant agreement
n$^{\rm o}$ 656370.  This work is supported in part by the Spanish
MINECO grant AYA2015-66899-C2-2-P, the programme PROMETEOII-2014-069
(Generalitat Valenciana), and by the NewCompstar COST action MP1304.

\appendix
\section{Wave equation coefficients} \label{sec:wvcoef}

We write here the coefficients of the wave equation~(\ref{eq:wv}):  
\begin{align}
 A_{1} &= \check  \mu \re ^{-2 \lambda} + \frac{1}{4 \pi} \left( B^{r} \right) ^{2}    \, , \\
 A_{2} & = \frac{ \check  \mu}{r^2}   +  \frac{1}{4 \pi}\left(  B^{\theta} \right)^2 \, , \\
 A_{3}  & =  \frac{1}{2 \pi} B^{r} B^{\theta }  \, , \\
 A_{4} & = \re ^{-2 \lambda}  \frac{d  \check  \mu}{d r}    + \left[   \frac{d}{dr} \hspace{-0.08cm} \left( \nu - \lambda \right) 
 + \frac{4}{r}  \right] \re ^{-2 \lambda}      \check  \mu 
 +    \frac{1}{4 \pi} B^{\theta}  \frac{ \partial  B^{r} }{\partial \theta } \nn \\
  &
 +  \frac{1}{4 \pi} \left( \frac{d \nu }{dr}+ \frac{2}{r} \right) \left( B^{r}  \right) ^{2} 
 +    \frac{1}{4 \pi} \left[ 2 \cot \theta B^{\theta} + \frac{ \partial B^{r} }{\partial r } \right]  B^{r} 
  \, ,  \\
 A_{5} & =  \frac{\cot \theta}{r^2} \left[ 3 \check  \mu +    \frac{1}{2 \pi} \left(  B^{\theta}  \right)^2 \right]
+ \frac{1}{4 \pi}  \left[   \left(  \frac{d\nu}{dr} + \frac{2}{r} \right) B^{\theta} + \frac{ \partial B^{\theta} }{dr }   \right] B^{r}   \nn \\
& +  \frac{1}{4 \pi} B^{\theta} \frac{ \partial  B^{\theta} }{\partial \theta }  
\end{align}
In the limit of zero shear modulus $\mu = 0$, the quantities $A_{k}$ become the coefficients of the wave equation for the core's protons.

\section{Crustal modes} \label{sec:cmAp}

The relativistic equations for studying the crustal torsional modes of
a superfluid star have been already derived
by~\citet{2009CQGra..26o5016S}.  By using the harmonic vector
expansion the problem becomes a 1D eigenvalue problem. For
axisymmetric modes the Lagrangian displacement can be written as
following
\begin{equation}
 \xi_{\ch}^{\phi} = \sum_{l}  \frac{\hat \xi_l}{\sin \theta} 
\frac{ \partial P_l }{\partial \theta}  \re^{\i \omega_l t}  \, ,   \label{eq:exp}
\end{equation}
where $\omega_l$ is the mode eigenfrequency, $\hat \xi_l$ is function
of $r$, and $P_l$ is the Legendre polynomial.  The label $l$ denotes
the harmonic index. If we introduce the variable
expansion~(\ref{eq:exp}) into the perturbation equation~(\ref{eq:wv})
with zero magnetic field, after some algebra, we find the following
equation in the coordinate basis:
\begin{align}
& \check \mu  \frac{ \partial^2 \hat \xi_l }{\partial r^2}  = 
 \left[ -  \frac{d  \check  \mu}{d r}  
+ \left( \frac{d  \lambda}{d r} - \frac{d  \nu}{d r}   - \frac{4}{r} \right) \check \mu  \right]  \frac{ \partial \hat \xi_l }{\partial r}  \nn \\ 
& + \left[   \left(l+2\right) \left( l - 1 \right)  \frac{ \check \mu  }{r^2}  - \left(\veps + p \right) \chi  \re^{-2\nu}  \omega_l^2  \right] \re^{2\lambda} \hat \xi_l  \, .
\end{align}
This equation can be solved as an eigenvalue problem with boundary
conditions at the crust/core interface, $r=R_{cc}$, and the star's
surface, $r=R$.  In both these boundaries, the continuity of traction
leads to the following condition:
\begin{equation}
\frac{ \partial \hat \xi_l }{\partial r} = 0 \, . 
\end{equation}

\nocite*
\bibliographystyle{mn2e}


\label{lastpage}

\end{document}